\documentclass[twocolumn,showpacs,preprintnumbers,amsmath,amssymb,floatfix]{revtex4-1}
\usepackage{graphicx}
\usepackage{epstopdf}
\usepackage{afterpage}
\usepackage{amsmath}
\usepackage{dcolumn}
\usepackage{bm}

\begin{document}

\preprint{APS/123-QED}

\title{Enhancement of level-crossing resonances in rubidium atoms by frequency control of the exciting radiation field}

\author{M.~Auzinsh}%
\email{Marcis.Auzins@lu.lv}%
\author{A.~Berzins}%
\author{R.~Ferber}%
\author{F.~Gahbauer}%
\author{L.~Kalvans}%
\author{A.~Mozers}%
\author{A.~Spiss}%

\affiliation{Laser Centre, The University of Latvia, 19 Rainis
Boulevard, LV-1586 Riga, Latvia}%

\date{\today}

\begin{abstract}
We studied magneto-optical resonances caused by excited-state level crossings in a nonzero magnetic field. 
Experimental measurements were performed on the transitions of the $D_2$ line of rubidium. These measured signals were described by a theoretical model that takes into account all neighboring hyperfine transitions, the mixing of magnetic sublevels in an external magnetic field, the coherence properties of the exciting laser radiation, and the Doppler effect. Good agreement between the experimental measurements and the theoretical model could be achieved over a wide range of laser power densities. We further showed that the contrasts of the level-crossing peaks can be sensitive to changes in the frequency of the exciting laser radiation as small as several tens of megahertz when the hyperfine splitting of the exciting state is larger than the Doppler broadening. 
\end{abstract}
\maketitle

\section{\label{Intro:level1}Introduction}
\begin{figure*}[htbp]
	\centering
		\resizebox{8cm}{!}{\includegraphics{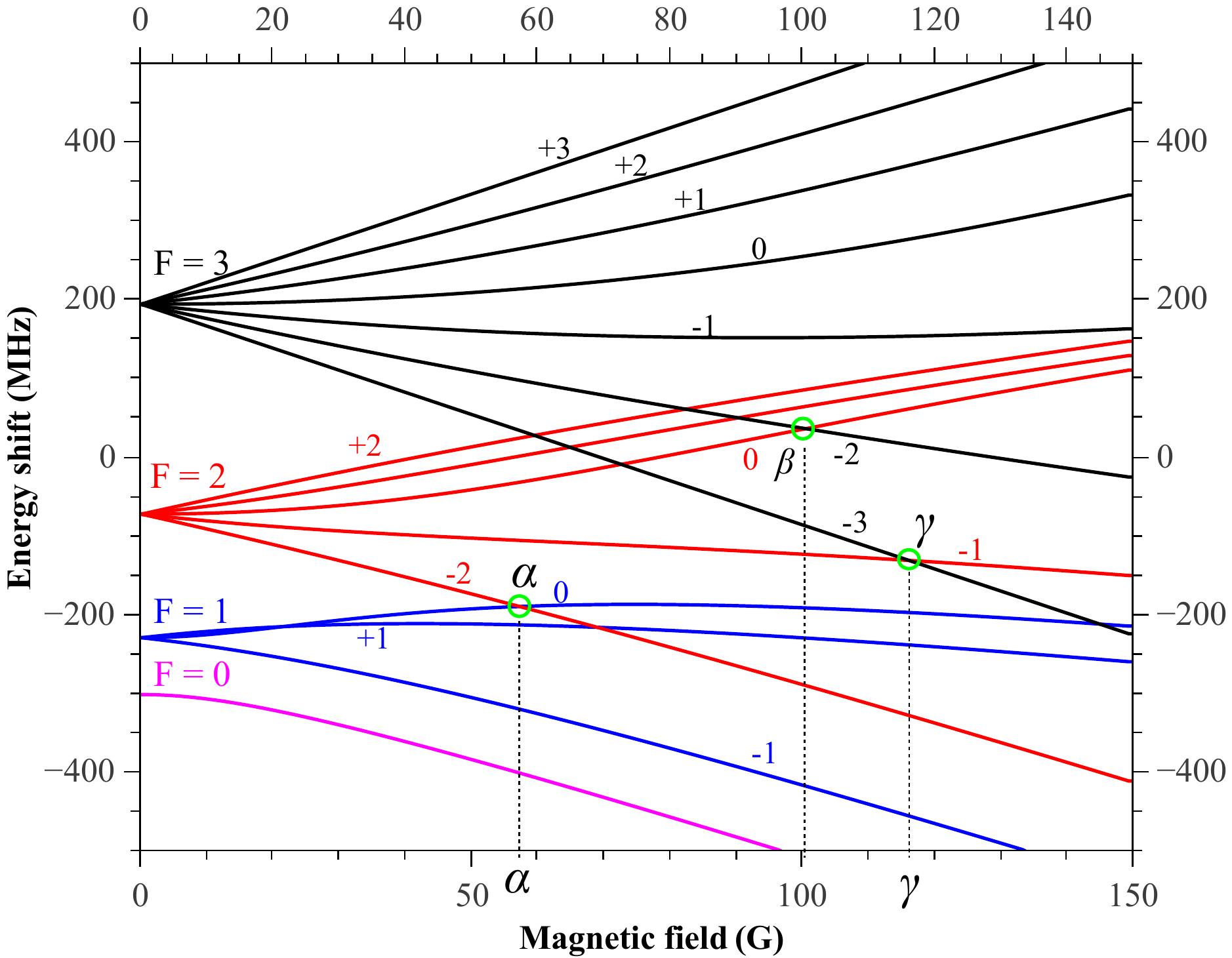}}
		\resizebox{8cm}{!}{\includegraphics{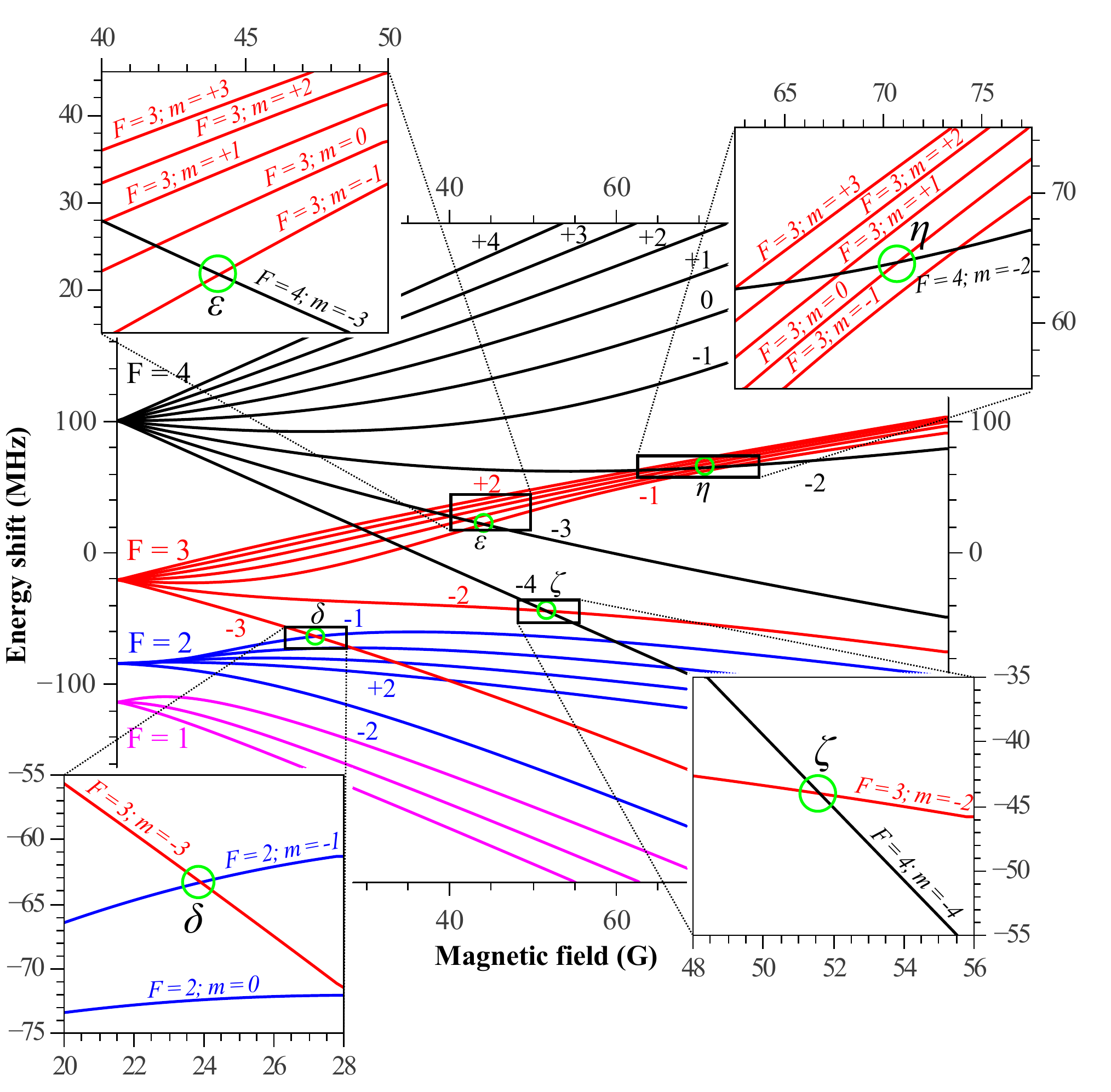}}
	\caption{\label{fig:levels} (Color online) Energy shifts as a function of magnetic field of excited-state hyperfine 
magnetic sublevels for $^{87}$Rb (left) and $^{85}$Rb (right). Zero energy corresponds to the excited-state fine structure level $5^2P_{3/2}$}
\end{figure*}

Level-crossing spectroscopy has long been used to study lifetimes of atomic states (using zero-field resonances) 
or atomic constants, such as the magnetic moments and fine and hyperfine constants (using nonzero-field resonances)~\cite{Arimondo:1977, Aleksandrov:1993}. 
The technique most often takes advantage of resonances in plots of the laser-induced fluorescence (LIF) in a particular direction 
with a given polarization as a function of the magnetic field. The resonances are related to the type of coherent excitation of magnetic sublevels that becomes possible when some of them, whose $z$-components $m_F$ of the total angular momentum $F$ differ by $\Delta m_F=q$, become degenerate at particular magnetic field values~\cite{Strumia}. For linearly polarized excitation $q=\pm 2$.
Such a degeneracy always occurs at zero magnetic field where all magnetic sublevels belonging to a particular hyperfine  level $F$ have the same energy. This case of zero-field level crossing is known as the Hanle effect, first observed by Hanle himself~\cite{Hanle:1924}. However, as can be seen in Fig.~\ref{fig:levels}, it also happens at certain nonzero magnetic field values that some magnetic sublevels from different hyperfine $F$ states can cross. If the requirement for coherently excited magnetic sublevels with a certain $\Delta m$ is fulfilled, one speaks of nonzero-magnetic-field level crossings.   
The coherent evolution of such systems can be described with the optical Bloch equations (OBEs) for the density matrix. 
However, in order to describe accurately real systems, it is necessary to take into account all neighboring hyperfine 
transitions, the magnetic-field-induced mixing of magnetic sublevels of identical $m$ that belong to different hyperfine levels, 
the Doppler profile, and the coherence properties of the radiation. Models with these characteristics have been developed over the years to describe 
zero-field resonances in the ground and excited states with great precision~\cite{Auzinsh:2012}. In this work, we show 
that nonzero level-crossing signals in magnetic fields can be described by a theoretical model over a wide range 
of magnetic fields to nearly experimental accuracy. Moreover, the model succeeds also at laser power densities for which the excitation is nonlinear and where the effects of optical pumping can be noted. 
We also show that, by carefully selecting and controlling the laser frequency with a precision of several tens of megahertz, it is possible to
obtain resonances with greater contrast, which can be useful in applications such as measurements of magnetic fields in 
the range of tens of Gauss or determination of hyperfine constants in states for which they are not yet known. The ability to describe level-crossing signals precisely can be useful for determining atomic constants especially in situations where the large number of crossing points washes out individual resonances. 
  
The first theory of level-crossing signals was given by Breit in 1933~\cite{Breit:1933}, and the first application was to measure the 
fine structure splitting between the helium $P$ states~\cite{Colegrove:1959}. 
These measurements were described in terms of Breit's formalism by Franken in 1961~\cite{Franken:1961}. 
Since then, these signals were used extensively for a time to make 
measurements of the fine and hyperfine constants in atoms. For example, the technique was used to obtain hyperfine constants in rubidium~\cite{Bucka:1966} and cesium~\cite{Svanberg:1969} (see Ref.~\cite{Arimondo:1977} for a review of many results). Theoretical models of the ground-state Hanle effect were used by Picqu\'e in 1978~\cite{Picque:1978}. Over time, these models became more and more sophisticated as different effects were included~\cite{Andreeva:2002}. Precise analytical models are also possible~\cite{Weis:2012}, 
but only for lower laser power in the linear regime. 

The present work revisits an earlier study published 
in 2003~\cite{Alnis:2003}, which offered only a model that was limited to the cycling transitions in the limit of weak excitation and thus could provide only a qualitative description of the experimental signals. We now show that the signals can be described very precisely even in the case of strong, nonlinear excitation with a model based on the optical Bloch equations and valid also for nonlinear excitation. Our model takes into account possible contributions to the transition probabilities from all neighboring hyperfine transitions, the effects of Doppler broadening, the splitting of the hyperfine levels in the magnetic field, and the coherence properties of the exciting laser radiation~\cite{Blushs:2004}. This model had been widely applied to zero-field resonances in the ground state~\cite{Auzinsh:2012} and achieved 
good agreement between experimentally measured and calculated curves. The experimental parameters, in particular the 
laser frequency, were carefully controlled during 
the measurements in order to allow precise comparison with theory. In addition, experimental measurements and theoretical 
calculations were used to investigate the influence of the laser frequency on the relative contrasts of the level-crossing 
peaks at nonzero field values.

\section{\label{Experiment:level1}Experiment}
Figure~\ref{fig:levels} shows the relative energies of the excited state magnetic sublevels as a function of the magnetic field. Each curve corresponds to a particular 
value $m_F$ of the projection of the total angular momentum $F$ on the $z$-axis. As we used linearly polarized exciting radiation, coherences can be formed around the crossing points with $\Delta m=\pm 2$, which are circled and labeled by small Greek letters. During the experiments, the laser frequency detuning is determined as the frequency difference between a particular ground-state hyperfine level and the excited-state fine structure level. 

\begin{figure}[htbp]
	\centering
		\resizebox{8cm}{!}{\includegraphics{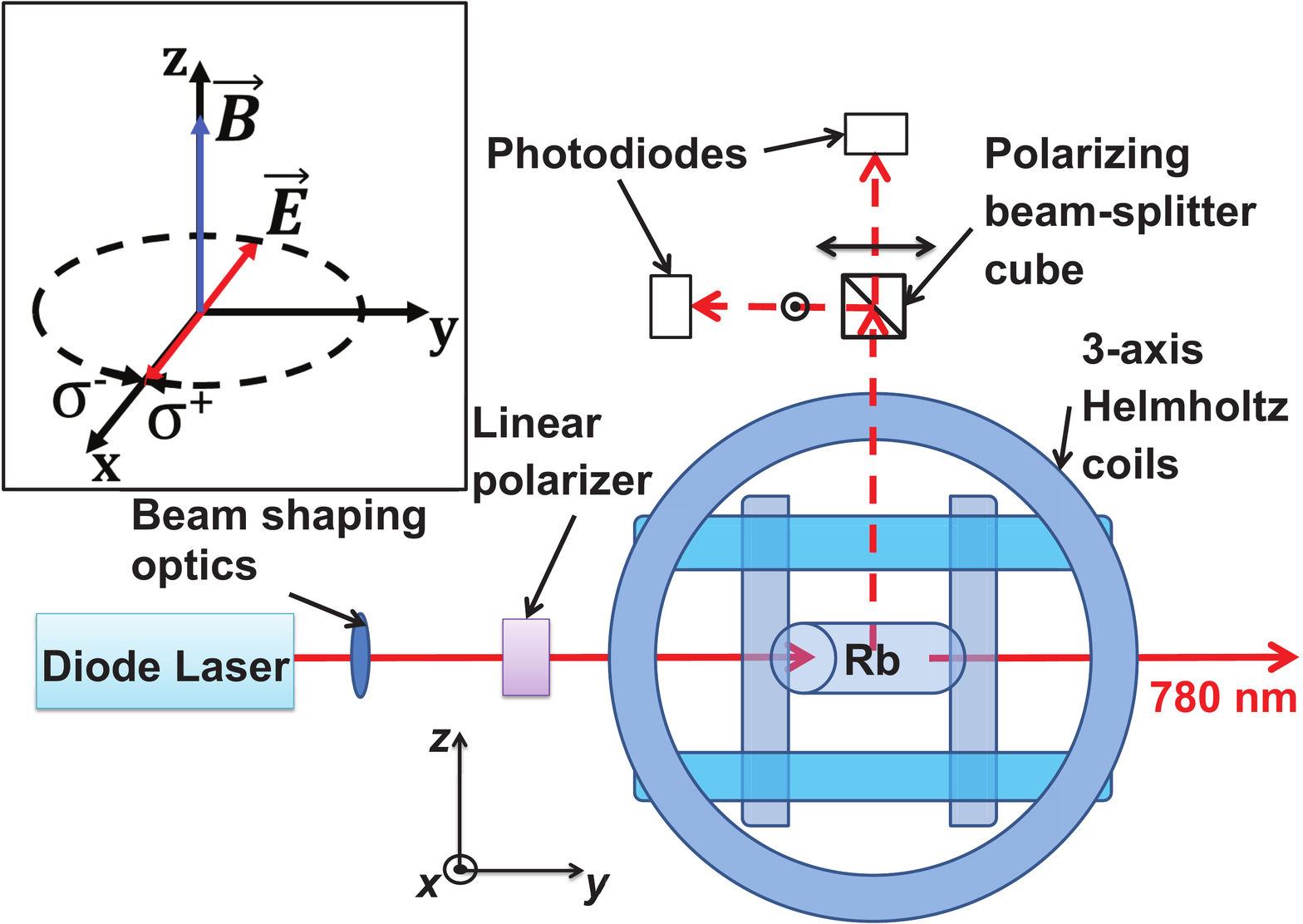}}
	\caption{\label{fig:geometry} (Color online) Schematic drawing of the experimental setup.
}
\end{figure}

In this experiment natural atomic rubidium, confined in a Pyrex cell with optical quality windows (25 mm long and 25 mm in diameter), was placed in the middle of a three-axis set of Helmholtz coils. Two pairs of coils were used to compensate the Earth's magnetic field, while the magnetic field was scanned along the third ($z$) axis. As the magnetic field was scanned through a triangular pattern with a frequency of 0.02 Hz, fluorescence spectra were acquired. The laser wavelength was determined by means of a saturated absorption spectroscopy setup in conjunction with a WS-7 wavemeter from HighFinesse. It was monitored during the scan with the wavemeter, and adjustments were made if necessary. Using a bipolar Kepco BOP-50-8-M or an Agilent N5770A power supply,  magnetic fields of up to 120 G could be achieved. Laser radiation from an external cavity diode laser passed through a chopper and entered the cell with its propagation vector and electric field vector both perpendicular to the scanning magnetic field (see Fig.~\ref{fig:geometry}). The temperature of the laser box and the diode were stabilized by Thorlabs TED200 temperature controllers and the current was controlled by a Thorlabs LDC205B current controller. The diameter of the beam was 1.6 mm as measured by a Thorlabs BP104-VIS beam profiler. The beam width was defined as the full width at half maximum of the Gaussian intensity profile. By means of a polarization rotator followed by a linear polarizer, laser power values from 20 $\mu$W to 320 $\mu$W could be achieved, which translated into laser power densities of 1 mW/cm$^2$ to 16 mW/cm$^2$. 
The LIF of two mutually perpendicular components (one parallel and the other perpendicular to the exciting electric field vector) was detected by two Thorlabs FDS100 photodiodes located behind a polarizing beam-splitter.  The signals were amplified by a transimpedance amplifier followed by a lock-in amplifier and recorded using an Agilent DSO5014A oscilloscope, which also averaged the signals. 

\section{\label{Theory:level1}Theoretical Model}
We describe the atomic system via its quantum density matrix $\rho$, which is written in the basis of Zeeman sublevels for 
the hyperfine structure of the $D_2$ transition of atomic Rubidium: $\vert\xi_i,F_i,m_{Fi}\rangle$,  where $F_i$ denotes 
the quantum number of the total atomic angular momentum in either the ground ($i = g$) or the excited ($i = e$) state, $m_{Fi}$ 
refers to the respective magnetic quantum number, and $\xi_i$ represents all other quantum numbers that remain unchanged 
within the $D_2$ line. The time evolution of the density matrix $\rho$ is governed by the optical Bloch equations~\cite{Stenholm:2005}:
\begin{equation}\label{eq:obe}
	i\hbar\frac{\partial \rho}{\partial t} = \left[\hat H,\rho\right] + i\hbar\hat R\rho,
\end{equation}
where $\hat H$ denotes the full Hamilton operator of the system and $\hat R$ is the relaxation operator. The full 
Hamiltonian can be expressed in terms of the unperturbed atomic Hamiltonian $\hat H_0$ determined by the internal dynamics of the atom, 
a term $\hat H_B$ that describes the interaction with the external magnetic field, and a dipole interaction 
term $\hat V = -\mathbf{\hat d}\cdot\mathbf{E}(t)$:
\begin{equation}\label{eq:ham}
	\hat H = \hat H_0 + \hat H_B + \hat V.
\end{equation}
As indicated, the interaction with the electromagnetic field is treated in the dipole approximation~\cite{optically-polarized-atoms}. The magnetic 
interaction Hamiltonian can be written as
\begin{equation}\label{eq:mag-ham}
	\hat H_B = \frac{\mu_B}{\hbar}(g_J\mathbf{J}+g_I\mathbf{I})\cdot \mathbf{B},
\end{equation}
where $\mu_B$ is the Bohr magneton, $\mathbf{J}$ and $\mathbf{I}$ are the total electronic angular momentum and spin of the 
atomic nucleus and $g_J$, $g_I$ are the respective Land\'{e} factors. The interaction Hamiltonian \eqref{eq:mag-ham} consists of
the interaction matrices for fixed projection $m_F = m_J + m_I$ of the angular momenta on the quantization axis, 
which can be written in terms of Wigner 3$j$ symbols~\cite{optically-polarized-atoms}. Solving the eigenvalue problem for these 
matrices yields the energy structure shown in Fig.~\ref{fig:levels}, where $\mathbf{J}$ and $\mathbf{I}$ make up the total atomic angular momentum 
$\mathbf{F} = \mathbf{J} + \mathbf{I}$. The components ($c_k$) of the eigenvectors are used as 
mixing coefficients for the chosen basis. The mixed atomic states have to be rewritten in the presence of the magnetic field as
\begin{equation}\label{eq:mixing}
	\vert \xi,F,m\rangle = \sum\limits_k c_k\vert \xi,F_k,m\rangle.
\end{equation}

The density matrix $\rho$ can be divided into four parts: $\rho_{gg}$ and $\rho_{ee}$ are quadrants aligned along the 
main diagonal and are called Zeeman coherences, $\rho_{ge}$ and $\rho_{eg}$ are called optical coherences. 
In order to describe the atomic fluorescence one has to know the elements $\rho_{ee}$. The atomic dipole interaction with 
a classically oscillating electric field characterized by a stochastic phase that gives raise to some spectral width $\Delta\omega$ 
of the frequency is considered. The rotating wave approximation~\cite{Allen:1975} and the decorrelation of and averaging over the 
stochastic phase~\cite{Kampen:1976} allows adiabatic elimination of the optical coherences~\cite{Blushs:2004}, which results in rate equations for the Zeeman coherences:
\begin{subequations} \label{eq:zc}
\allowdisplaybreaks
\begin{align}
\frac{\partial \rho_{g_ig_j}}{\partial t} =& \left(\Xi_{g_ie_m}
 + \Xi_{e_kg_j}^{\ast}\right)\sum_{e_k,e_m}d_{g_ie_k}^\ast d_{e_mg_j}\rho_{e_ke_m} - \nonumber\\ 
-&\sum_{e_k,g_m}\Big(\Xi_{e_kg_j}^{\ast}d_{g_ie_k}^\ast d_{e_kg_m}\rho_{g_mg_j} + \nonumber\\ 
+& \Xi_{g_ie_k}d_{g_me_k}^\ast d_{e_kg_j}\rho_{g_ig_m}\Big) - \nonumber\\
-&i \omega_{g_ig_j}\rho_{g_ig_j} + \sum_{e_ke_l}\Gamma_{g_ig_j}^{e_ke_l}\rho_{e_ke_l} - \nonumber\\
-& \gamma\rho_{g_ig_j} + \lambda\delta(g_i,g_j) \label{eq:zcgg} \\
\frac{\partial \rho_{e_ie_j}}{\partial t} =& \left(\Xi_{e_ig_m}^\ast + \Xi_{g_ke_j}\right)\sum_{g_k,g_m}d_{e_ig_k} 
d_{g_me_j}^\ast\rho_{g_kg_m} - \nonumber\\
-& \sum_{g_k,e_m}\Big(\Xi_{g_ke_j}d_{e_ig_k} d_{g_ke_m}^\ast\rho_{e_me_j} + \nonumber\\ 
+& \Xi_{e_ig_k}^\ast d_{e_mg_k} d_{g_ke_j}^\ast\rho_{e_ie_m}\Big) - \nonumber\\
-& i \omega_{e_ie_j}\rho_{e_ie_j} - (\Gamma + \gamma)\rho_{e_ie_j}. \label{eq:zcee}
\end{align}
\end{subequations}
The following terms are used in Eq.~\eqref{eq:zc}: $\Xi_{ij}$ describes the interaction strength between the atom and the laser radiation and is defined below, 
$d_{ij}$ is the dipole transition matrix element that can be obtained from the reduced matrix element by means of the 
Wigner-Eckart theorem~\cite{optically-polarized-atoms}, 
$\omega_{ij}$ is the energy difference between levels $\vert i\rangle$ and $\vert j\rangle$, $\Gamma_{g_ig_j}^{e_ke_l}$ 
describes coherence transfer to the ground state via spontaneous emission, $\gamma$ is the transit relaxation rate at 
which the atoms leave the interaction region, $\Gamma$ is the rate of the spontaneous transitions, and $\lambda$ describes 
rate at which ``fresh'' atoms enter the interaction region. We assume that atoms entering the interaction region 
are completely depolarized and that the atomic equilibrium density outside the interaction region is normalized to unity; thus,  $\lambda = \gamma$.

The interaction strength $\Xi_{ij}$ is given by
\begin{equation}\label{eq:xi}
	\Xi_{ij} = \frac{\vert\varepsilon_{\bar\omega}\vert^2}{\hbar^2}\frac{1}{\frac{\Gamma+\gamma+\Delta\omega}{2}+\dot\imath\left(\bar\omega-\mathbf{k}_{\bar\omega}\mathbf{v}
	+\omega_{ij}\right)},
\end{equation}
where $\varepsilon_{\bar\omega}$ is the amplitude of the oscillating electric field, $\bar\omega$ denotes the central 
frequency of the laser radiation, $\Delta \omega$ is the laser linewidth, and $\mathbf{k_{\bar\omega}v}$ is the Doppler shift of the atomic transition for an atom moving with velocity $\mathbf{v}$.
In the numerical simulations we define the Rabi frequency in the following way:
\begin{equation}\label{eq:Rabi}
	\Omega_R = \frac{\vert\varepsilon_{\bar\omega}\vert\cdot\vert\vert d_J\vert\vert}{\hbar},
\end{equation}
where $\vert\vert d_J\vert\vert$ is the reduced dipole element of the $D_2$ transition whose value can be found 
in~\cite{optically-polarized-atoms}. The Rabi frequency is used as a parameter that corresponds to the experimentally measurable excitation 
laser power density:
\begin{equation}\label{eq:Rabi-power}
	I = k_{Rabi}\Omega_R^2.
\end{equation}
The proportionality coefficient $k_{Rabi}$ can be estimated from Eq. \eqref{eq:Rabi}. However, the precise value must be determined from the best fit to the experimental data because the laser power density is not constant over the beam profile.

In this study we numerically solved equations \eqref{eq:zc} for steady state excitation conditions $\left(\frac{\partial\rho_{g_ig_j}}{\partial t} = \frac{\partial\rho_{e_ie_j}}{\partial t} = 0\right)$ to obtain the Zeeman coherences $\rho_{gg}$ and $\rho_{ee}$. From this point it was straightforward to obtain the fluorescence for some particular polarization component defined by $\mathbf{e}$:
\begin{equation}\label{eq:fluo}
	I_{fl}(\mathbf{e}) = \tilde I_0 \sum_{g_i,e_j,e_k} d_{g_ie_j}^{(ob)\ast}d_{e_kg_i}^{(ob)}\rho_{e_je_k},
\end{equation}
where $d_{ij}^{(ob)}$ are the dipole transition matrix elements for the chosen observation component.

In order to take into account the classical movement of atoms with velocity $\mathbf{v}$ we performed a numerical integration 
of the fluorescence signal over the frequency distribution in the Doppler profile.

\section{\label{Results:level1}Results and Discussion}
\begin{figure*}[htbp]
	\centering
		\resizebox{8cm}{!}{\includegraphics{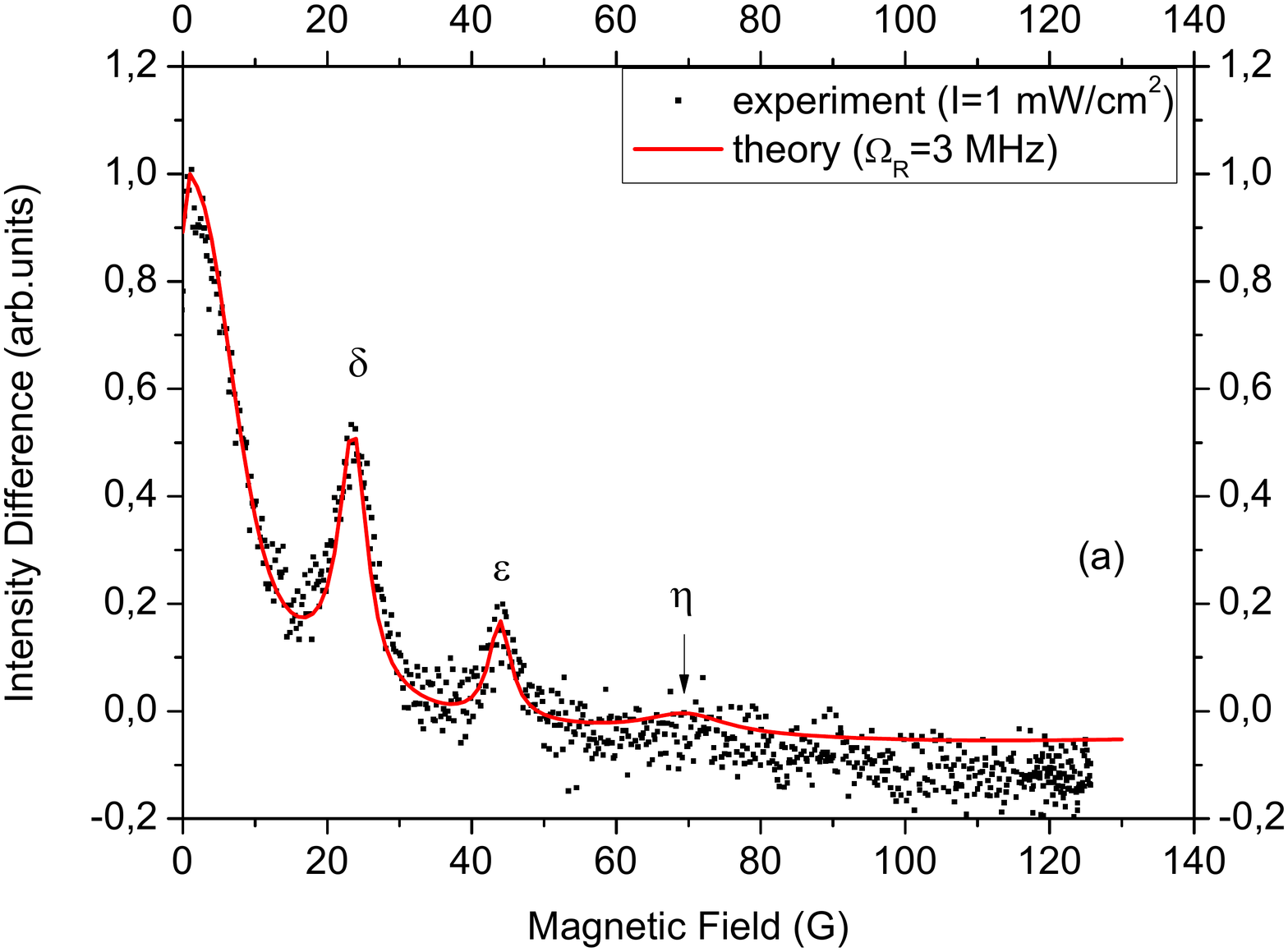}}
                \resizebox{8cm}{!}{\includegraphics{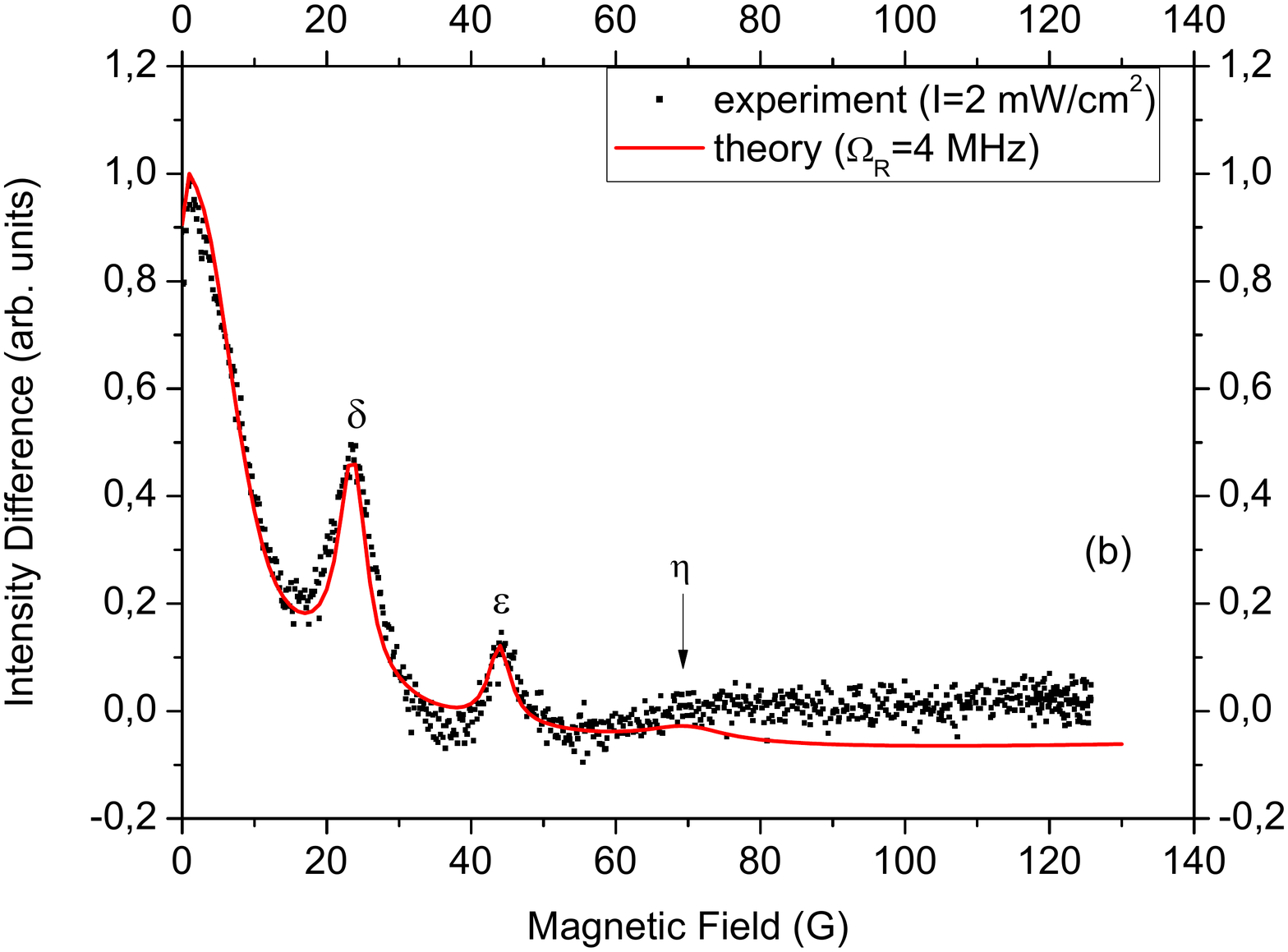}}
		\resizebox{8cm}{!}{\includegraphics{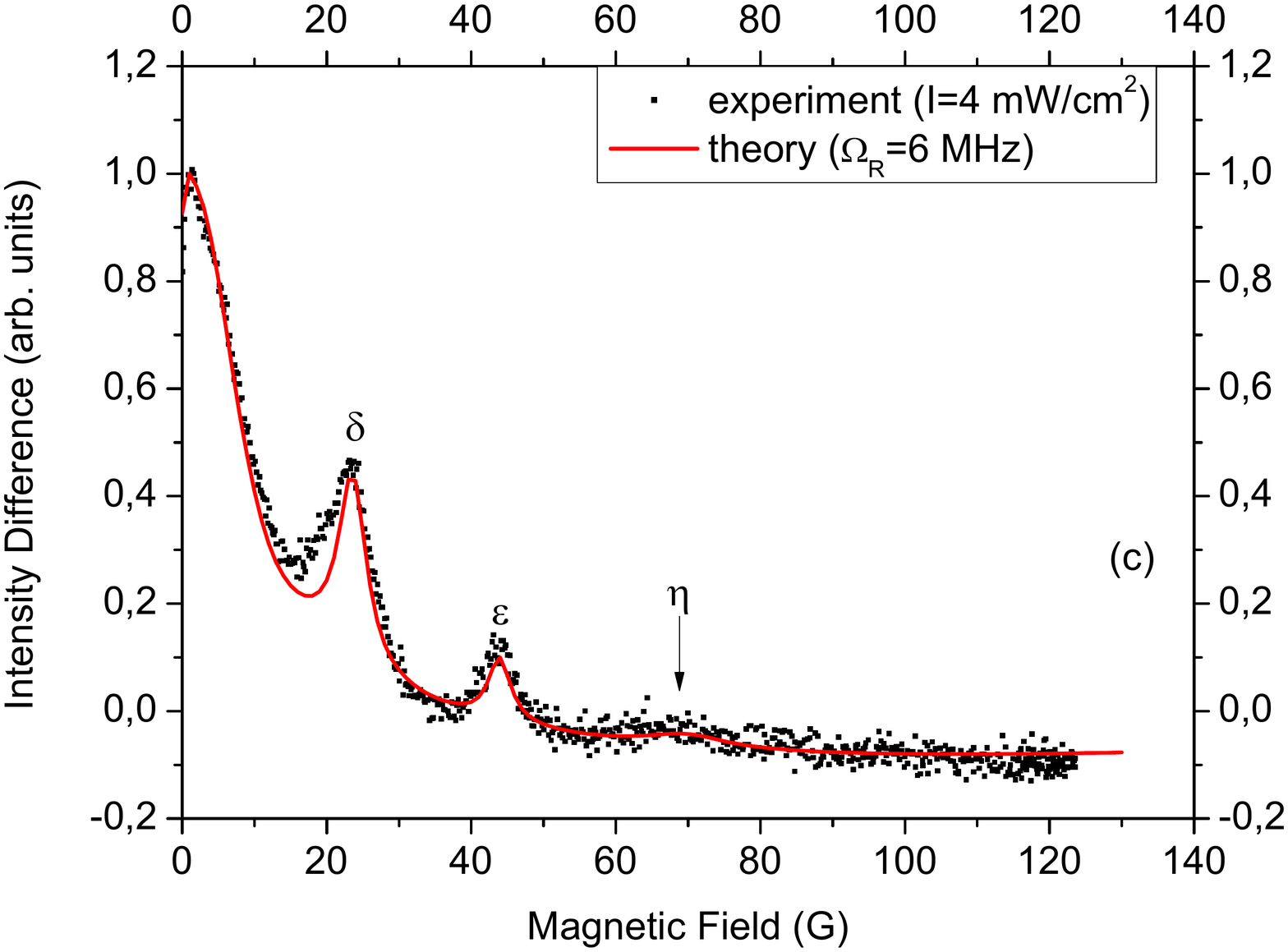}}
		\resizebox{8cm}{!}{\includegraphics{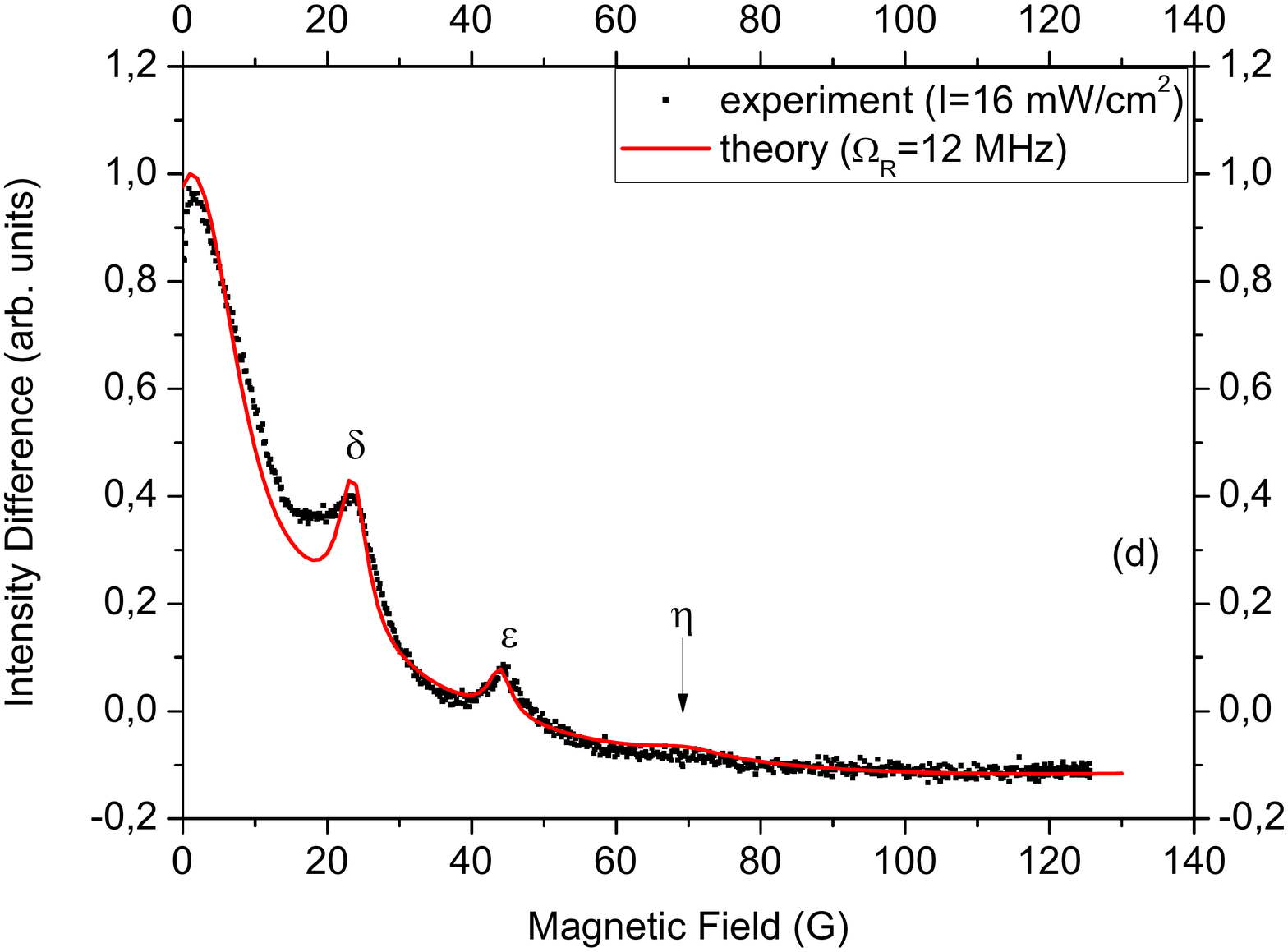}}
	\caption{\label{rb85-power} (Color online) Intensity difference ($I_{\perp}-I_{\parallel}$) versus magnetic field for the $F_g=2\longrightarrow F_e=1,2,3$ transitions of $^{85}$Rb. The different panels correspond to different laser power densities. Markers represent the results of 
experimental measurements, while the curves represent the results of theoretical calculations. 
}
\end{figure*}

As a first test of how well the model can describe non-zero level-crossing signals, we used the model to reproduce the 
experimental results of non-zero level crossing signals for the $F_g=2\longrightarrow F_e=1,2,3$ transitions of 
$^{85}$Rb. The results are shown in Fig.~\ref{rb85-power}. Measurements were carried out for a range of laser power 
densities $I$ from 1 mW/cm$^2$ to 16 mW/cm$^2$. The Rabi frequency $\Omega_R$ was assumed to vary as in Eq.~\eqref{eq:Rabi-power}, where the same value for $k_{Rabi}$ was used for all calculations in this paper. This value was chosen 
to provide the best overall agreement while being consistent with a rough estimation based on theoretical considerations (see Eq.~\eqref{eq:Rabi}). 
The other parameters that were optimized in a similar way were the laser linewidth $\Delta \omega$ and the proportionality constant between the laser beam 
width and the transit relaxation time. In each case only one constant value was used 
in all the theoretical calculations presented in this work. The agreement between theory and experiment was excellent at low laser power densities
and still quite good at $I=16$ mW/cm$^2$. At higher laser power densities the model becomes less accurate, because at higher power it is no longer 
sufficient to describe the transit relaxation by a single rate constant~\cite{Auzinsh:1983} as this model does.  
Nevertheless, the model seems adequate to predicting signals for level-crossing spectroscopy with $\Delta m=\pm2$.

\begin{figure*}[htbp]
	\centering
		\resizebox{8cm}{!}{\includegraphics{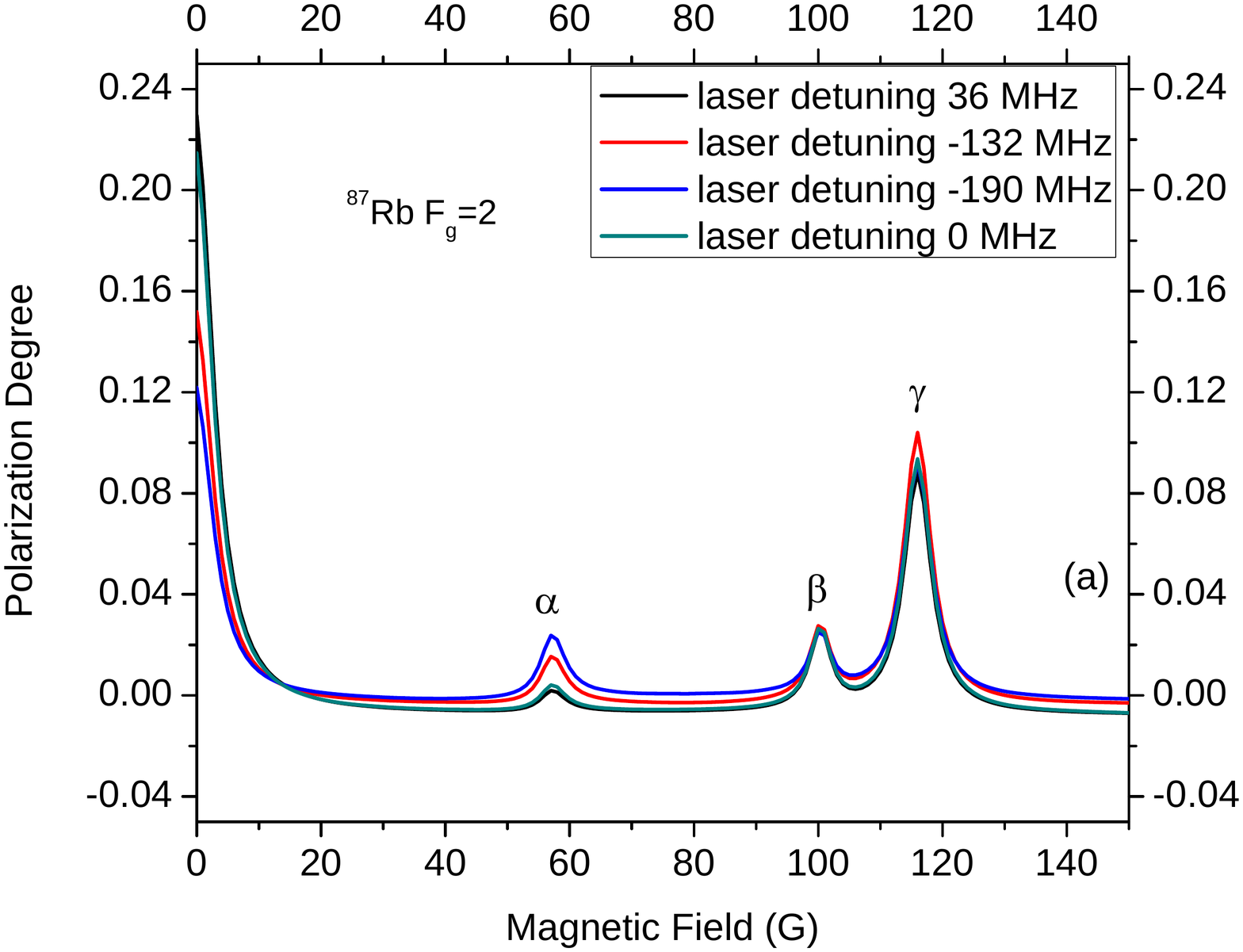}}
                \resizebox{8cm}{!}{\includegraphics{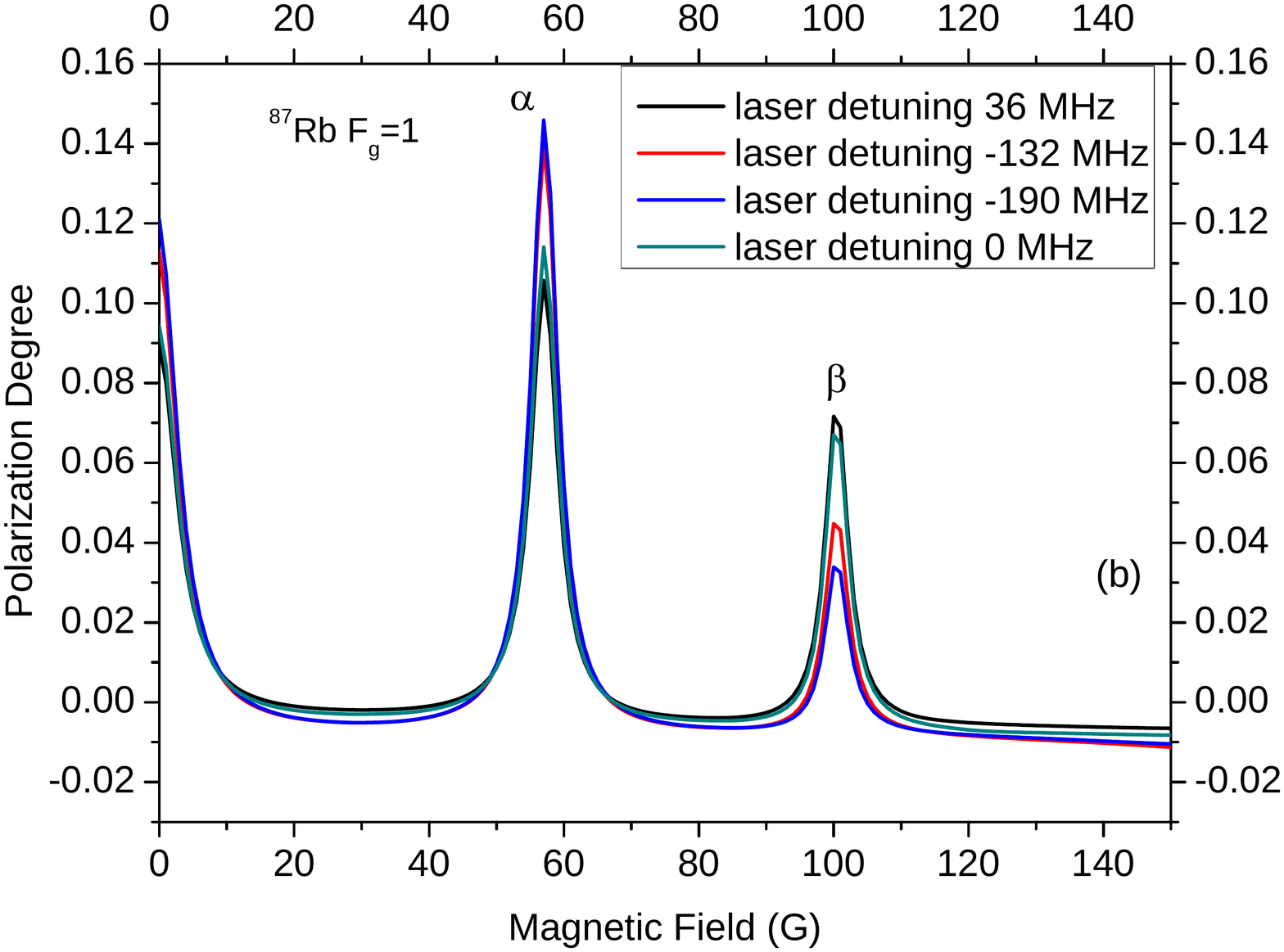}}
		\resizebox{8cm}{!}{\includegraphics{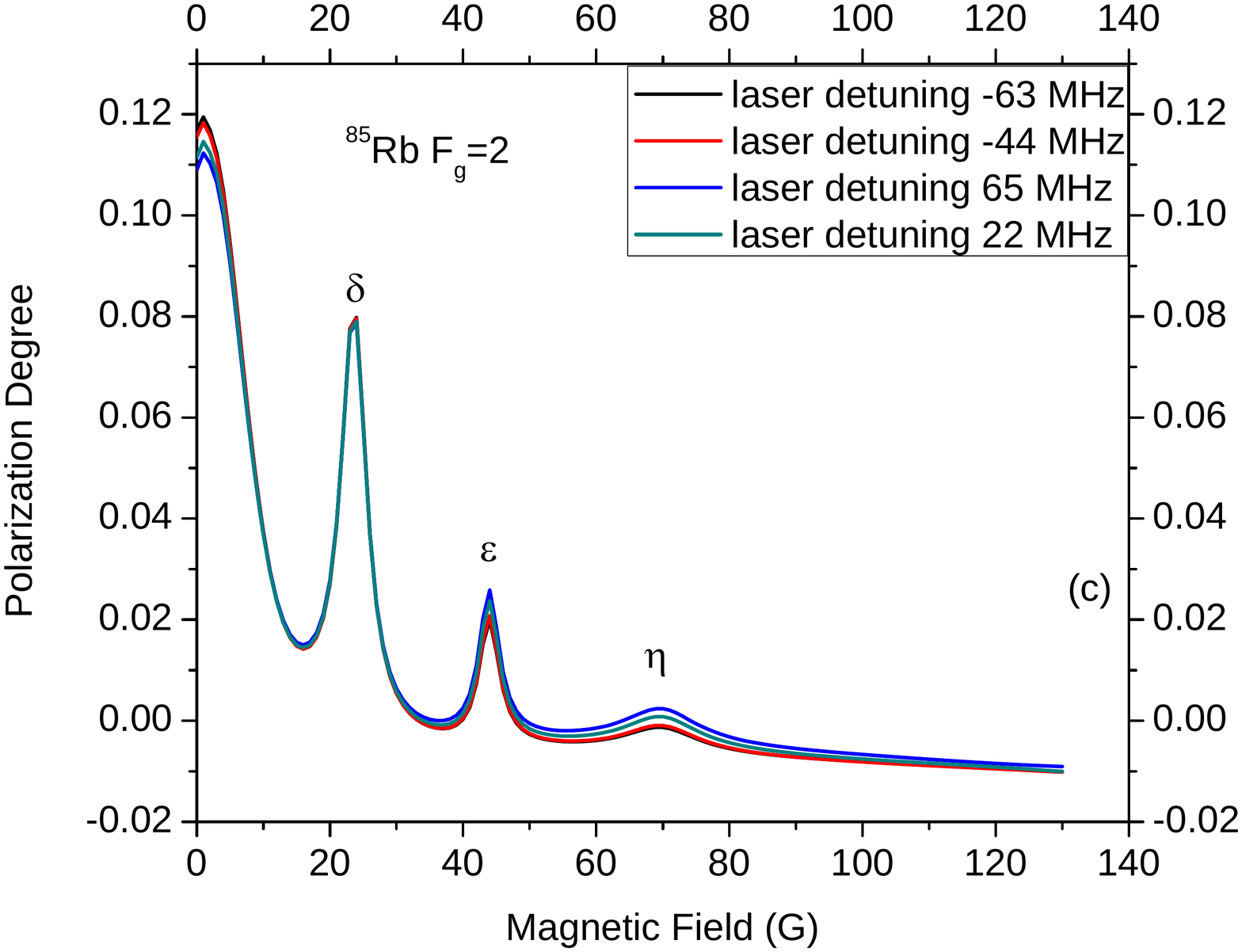}}
		\resizebox{8cm}{!}{\includegraphics{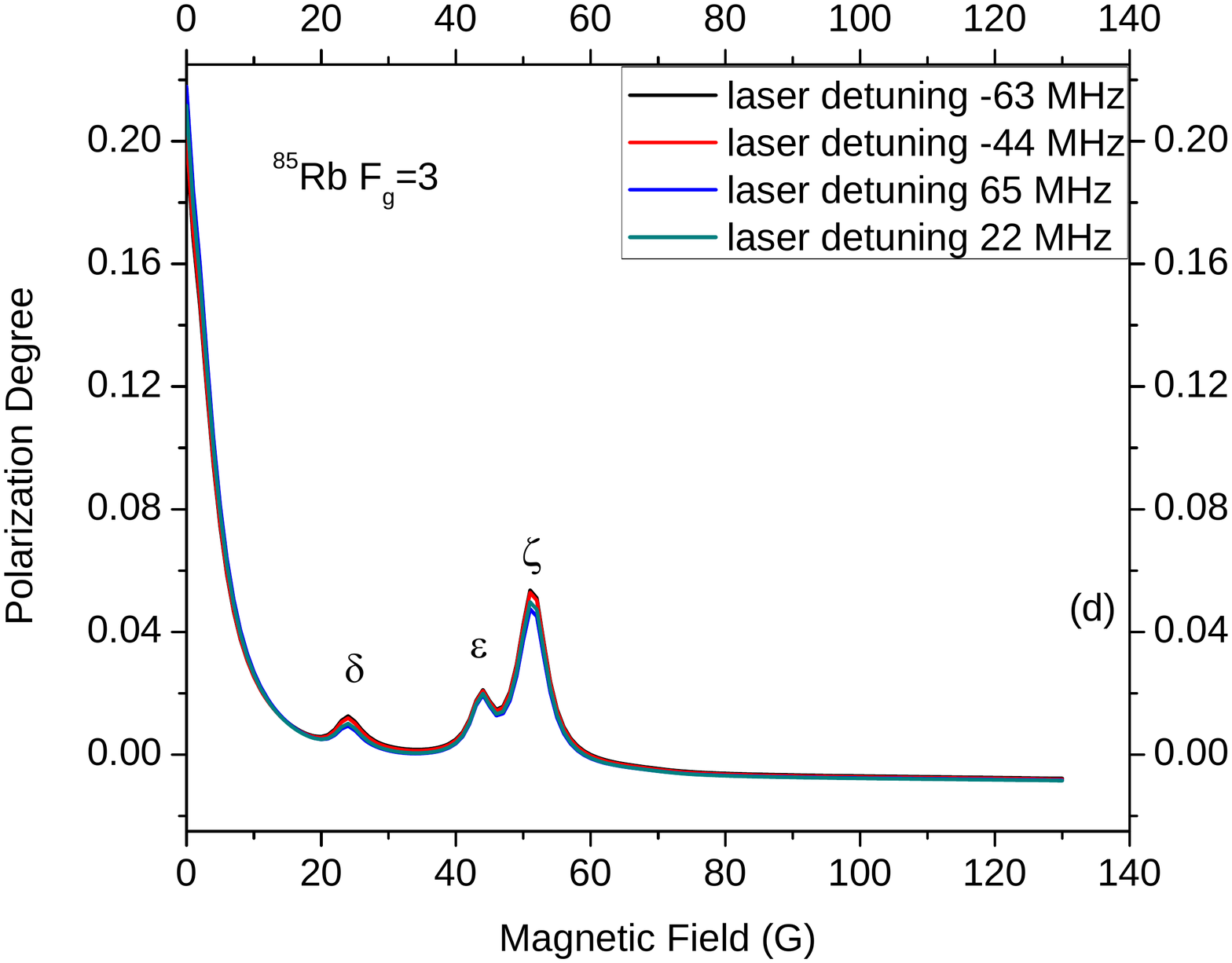}}
	\caption{\label{theory-detuning} (Color online)Theoretical calculations of $I_{\perp}-I_{\parallel}$ versus magnetic field for 
(a) the $F_g=2\longrightarrow F_e=1,2,3$ transition of $^{87}$Rb, (b) the $F_g=1\longrightarrow F_e=0,1,2$ transition of $^{87}$Rb, 
(c) the $F_g=2\longrightarrow F_e=1,2,3$ transition of $^{85}$Rb, and 
(d) the $F_g=3\longrightarrow F_e=2,3,4$ transition of $^{85}$Rb. 
The different curves in each figure correspond to different laser frequencies. The peaks are labeled by small Greek letters to indicate which crossing 
points gives rise to them (see Fig.~\ref{fig:levels}).
}
\end{figure*}

Next we turned our attention to the influence of the laser detuning on the level-crossing spectra. 
Figure~\ref{theory-detuning} presents the results of theoretical calculations that show how the relative contrasts of the 
nonzero-field level-crossing peaks can be influenced by the tuning of the exciting laser radiation. 
The detunings of the different curves plotted are measured with respect to the energy difference between the indicated ground-state hyperfine level and the excited-state fine structure level. First of all, it was 
apparent that significant variations in the relative contrast of the level-crossing peaks could be achieved only in $^{87}$Rb
[Fig.~\ref{theory-detuning}(a,b)].
The contrast of the peak at level crossing $\alpha$ could be increased by up to 
a factor of four when exciting from the $F_g$=2 ground state. When exciting from the $F_g$=1 ground state, the contrast of the peak that corresponds to
level crossing $\beta$ can be increased by almost a factor of two. 
In the case of $^{85}$Rb [Fig.~\ref{theory-detuning}(c,d)] Doppler broadening washes out any possible effects, 
because the hyperfine splitting of the excited state in this isotope is small relative to the Doppler width. The energy difference between the $F=1$ and $F=4$ states of $^{85}$Rb is slightly more than 200 MHz, whereas the full width at half maximum of the Doppler profile is around 500 MHz at room temperature. Although the change in laser frequency was smaller in absolute terms for the calculations with $^{85}$Rb than for the calculations with $^{87}$Rb, the change in detuning as a fraction of hyperfine splitting was roughly equal. Detuning the laser by a larger amount for $^{85}$Rb would result in significantly lower signals without affecting peak contrast significantly.   

\begin{figure*}[htbp]
	\centering
		\resizebox{\columnwidth}{!}{\includegraphics{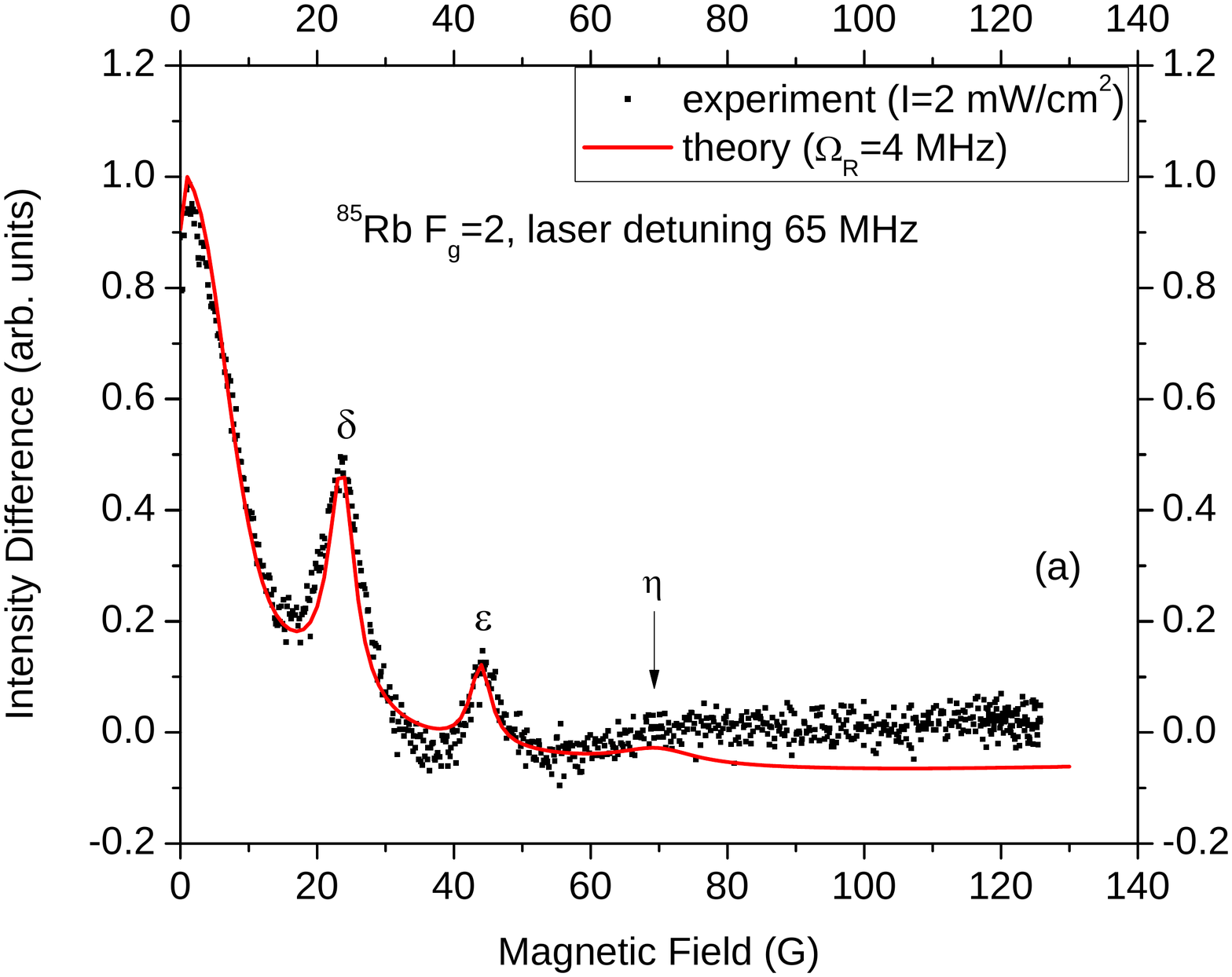}}
		\resizebox{\columnwidth}{!}{\includegraphics{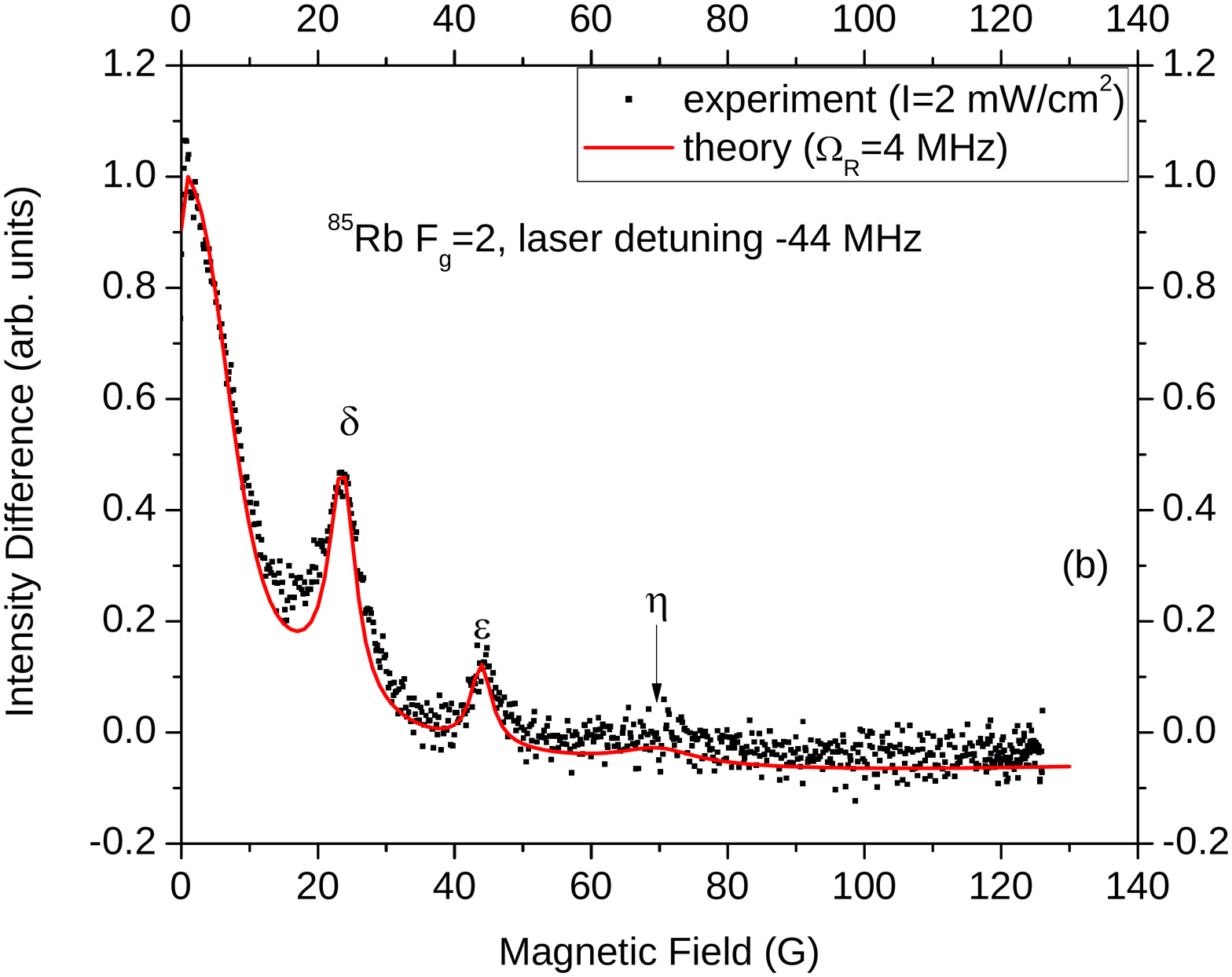}}
	\caption{\label{rb85-fg2}(Color online) Intensity difference ($I_{\perp}-I_{\parallel}$) versus magnetic field for the $F_g=2\longrightarrow F_e=1,2,3$ transitions of 
$^{85}Rb$ with the laser detuned from the $5^2S_{1/2} (F_g=2) \longrightarrow 5^2P_{3/2}$ transition by (a) 65 MHz and (b) $-44$ MHz. 
Markers show the results of an experiment with the laser power density $I=2$ mW/cm$^2$, while the solid line shows the results of a calculation with Rabi frequency $\Omega_R=4$ MHz.
}
\end{figure*}

\begin{figure*}[htbp]
	\centering
		\resizebox{\columnwidth}{!}{\includegraphics{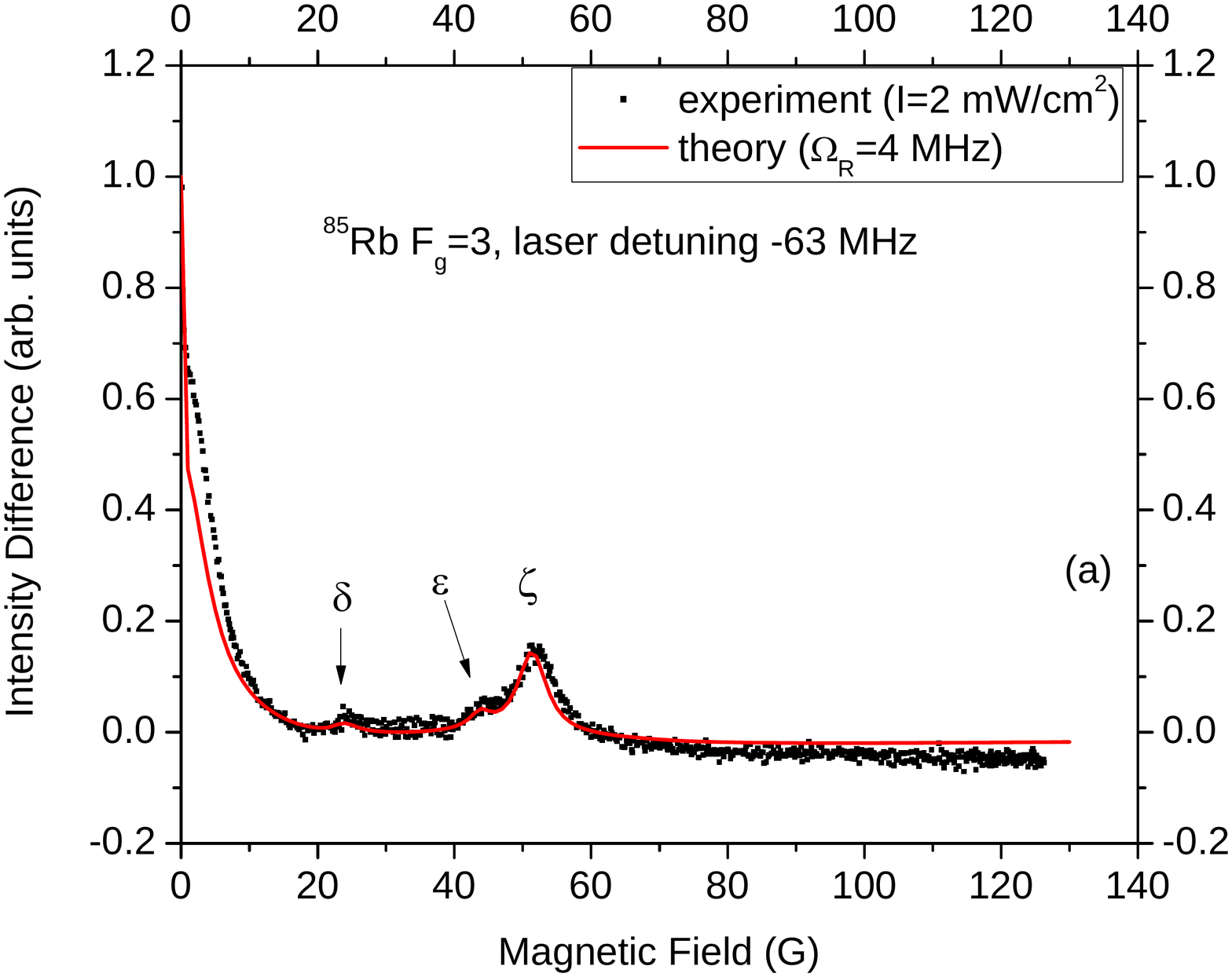}}
		\resizebox{\columnwidth}{!}{\includegraphics{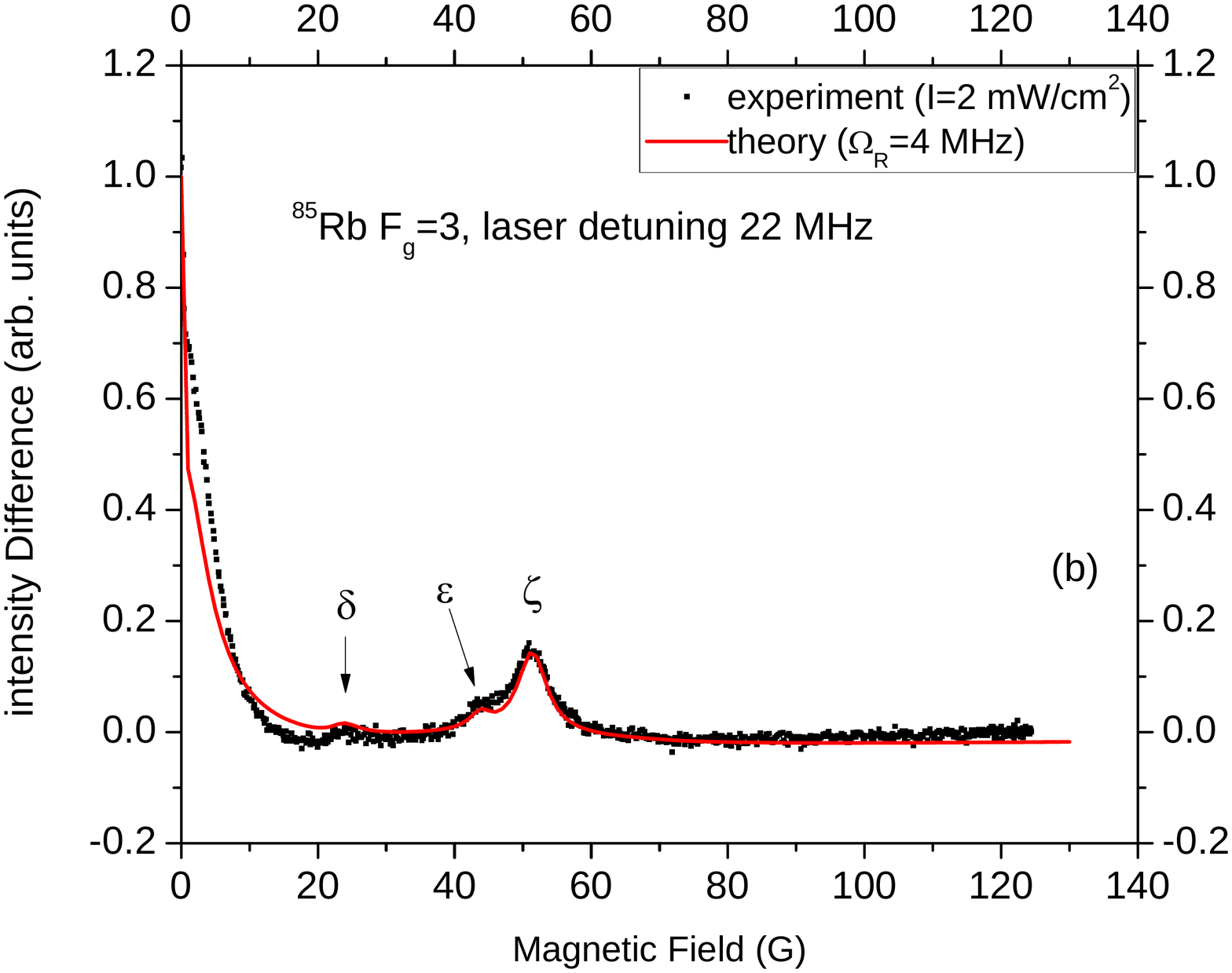}}
	\caption{\label{rb85-fg3} (Color online) Intensity difference ($I_{\perp}-I_{\parallel}$) versus magnetic field for the $F_g=3\longrightarrow F_e=2,3,4$ transitions of 
$^{85}$Rb with the laser detuned from the $5^2S_{1/2} (F_g=3) \longrightarrow 5^2P_{3/2}$ transition by (a) $-63$ MHz and by (b) $22$ MHz.   
Markers show the results of an experiment with the laser power density $I=2$ mW/cm$^2$, while the solid line shows the results of a calculation with Rabi frequency $\Omega_R=4$ MHz.
}
\end{figure*}

Next we compare experimental level-crossing curves obtained at different values of the laser detuning with the calculated curves 
for those same detuning values to see if the predicted changes in contrast can be observed. Figure~\ref{rb85-fg2}  shows results for $^{85}$Rb. 
Figure~\ref{rb85-fg2}(a) shows the results of an experiment in which the laser was detuned from the the exact energy difference between the ground state with $F_g=2$ and the $5^2P_{3/2}$ state by 65 MHz. The results shown in Fig.~\ref{rb85-fg2}(b) were obtained with a laser detuning of $-44$ MHz. Similarly, Fig.~\ref{rb85-fg3}(a) corresponds to a measurement with 
the laser detuning between the ground state of $^{85}$Rb with $F_g=3$ and the $5^2P_{3/2}$ state of $-63$ MHz, while Fig.~\ref{rb85-fg3}(b) corresponds to an experiment with a laser detuning of 22 MHz.
As expected, the relative contrasts of the level-crossing peaks are not very sensitive to detuning. 
We note that the theoretical model describes the experimental curves quite well. These transitions also illustrate the necessity of including magnetic sublevel mixing in the theoretical model. For example, the resonance at position $\epsilon$ (see Fig.~\ref{fig:levels}) involves a crossing of the sublevels labeled by $F_e=3,m_F=-1$ and $F_e=4,m_F=-3$. Notwithstanding the selection rules for $B=0$ ($\Delta F=0,\pm 1$), this level crossing can produce a resonance even when the excitation takes place from the ground state with
$F_g=2$. The reason is that $F$ ceases to be a good quantum number when the magnetic field is nonzero, and, at high values of the magnetic field, there is a sufficiently high probability of exciting the state labeled by $F_e=4,m_F=-3$ from $F_g=2$. However, the state with maximum projection of angular momentum ($F_e=4, m_F=4$) is not mixed. Thus, it is not possible to excite it from $F_g=2$, but only from $F_g=3$. These considerations are borne out by the theoretical and experimental results shown in Fig.~\ref{rb85-fg2} and Fig.~\ref{rb85-fg3}. A theoretical model, such as the one in~\cite{Alnis:2003}, that did not take into account magnetic sublevel mixing at high magnetic fields would not have been able to reproduce the resonance at position $\epsilon$ for an excitation  from $F_g=2$~\cite{Alnis:2003}. 

\begin{figure*}[htbp]
	\centering
		\resizebox{\columnwidth}{!}{\includegraphics{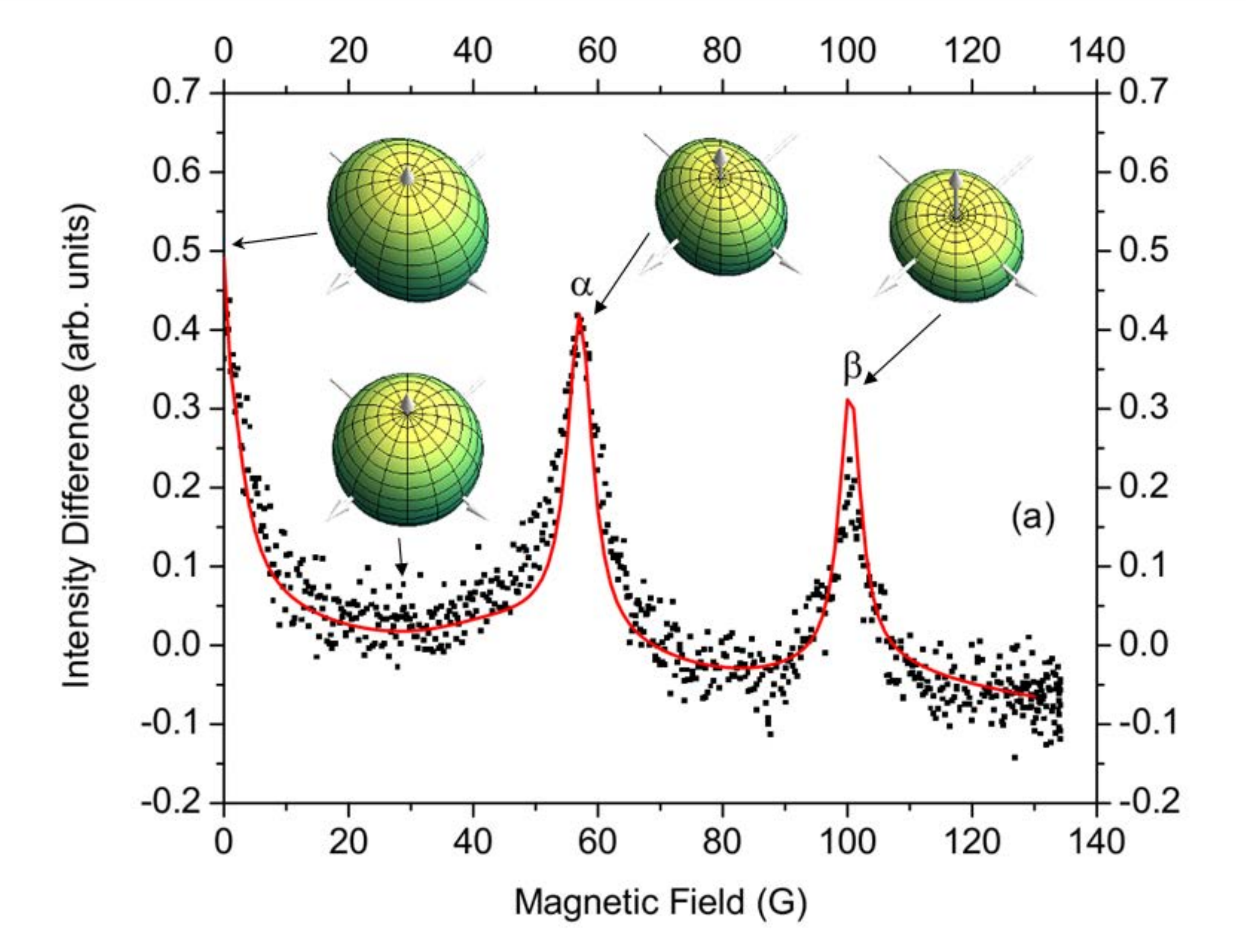}}
		\resizebox{\columnwidth}{!}{\includegraphics{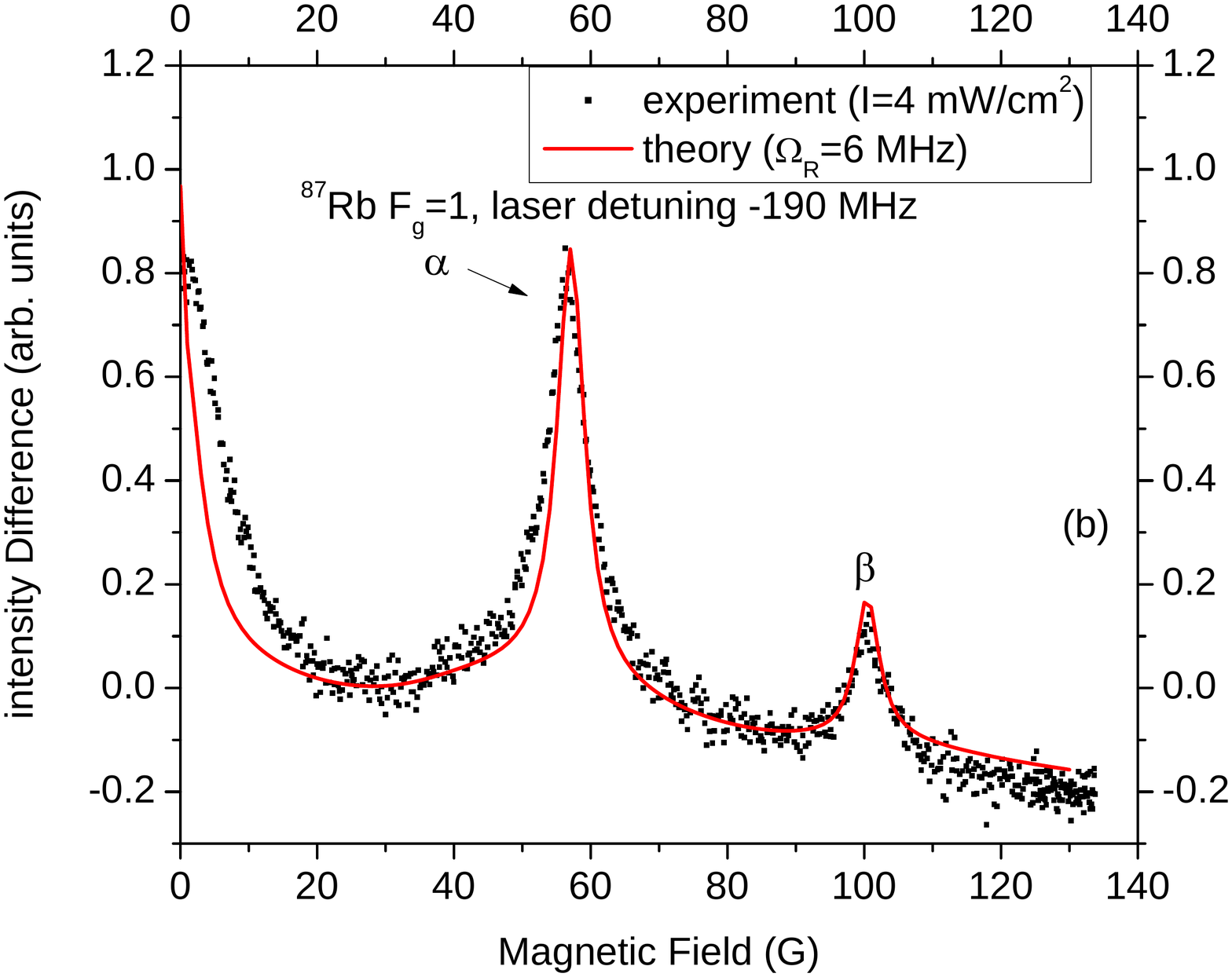}}
	\caption{\label{rb87-fg1} (Color online) Intensity difference ($I_{\perp}-I_{\parallel}$) versus magnetic field for the $F_g=1\longrightarrow F_e=0,1,2$ transitions of 
$^{87}$Rb with the laser detuned from the $5^2S_{1/2} (F_g=1) \longrightarrow 5^2P_{3/2}$ transition by (a) 36 MHz and by (b) $-190$ MHz. 
Markers show the results of an experiment with the laser power density $I=4$ mW/cm$^2$, while the solid line shows the results of a calculation with Rabi frequency $\Omega_R=6$ MHz.
}
\end{figure*}

\begin{figure*}[htbp]
	\centering
		\resizebox{\columnwidth}{!}{\includegraphics{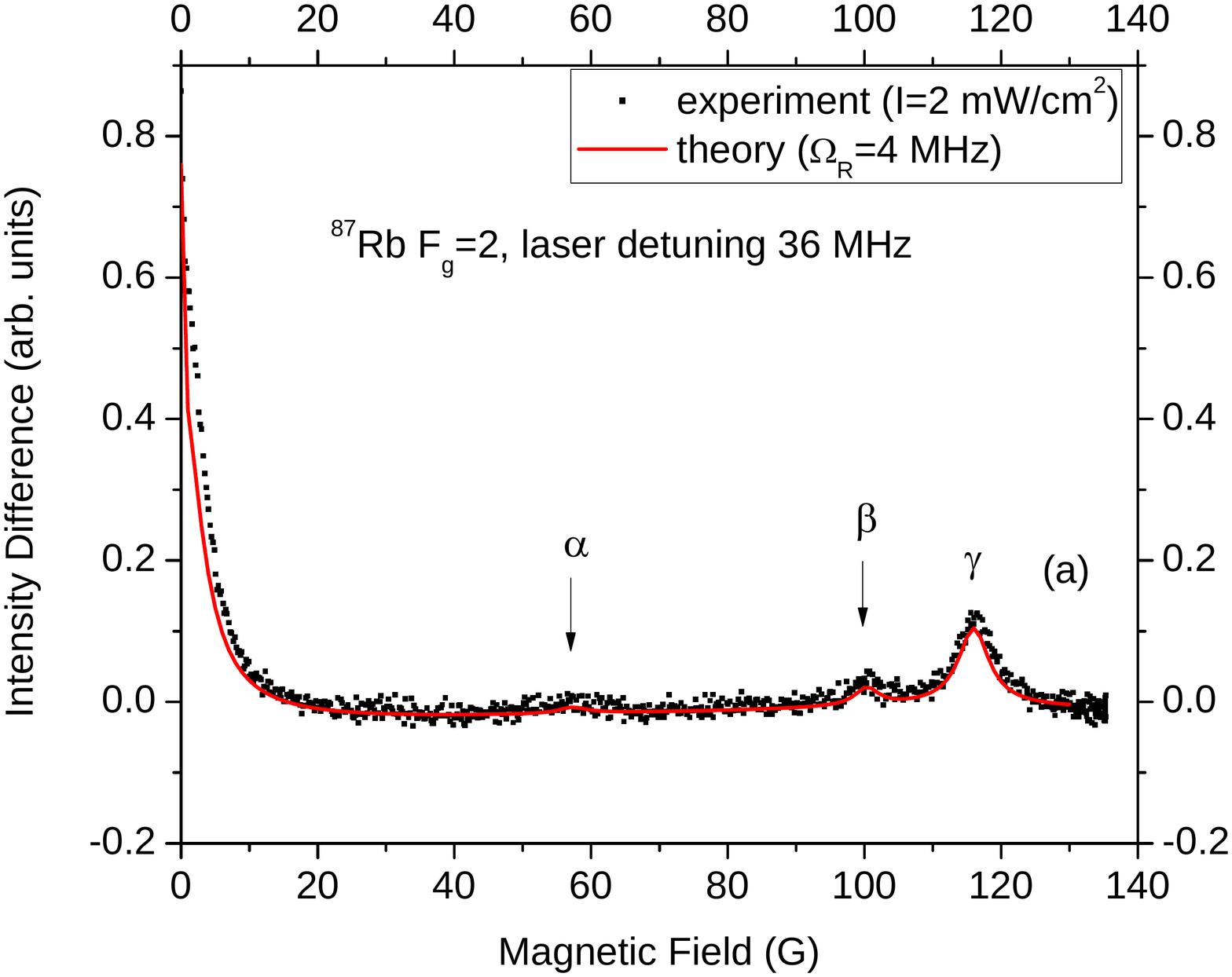}}
		\resizebox{\columnwidth}{!}{\includegraphics{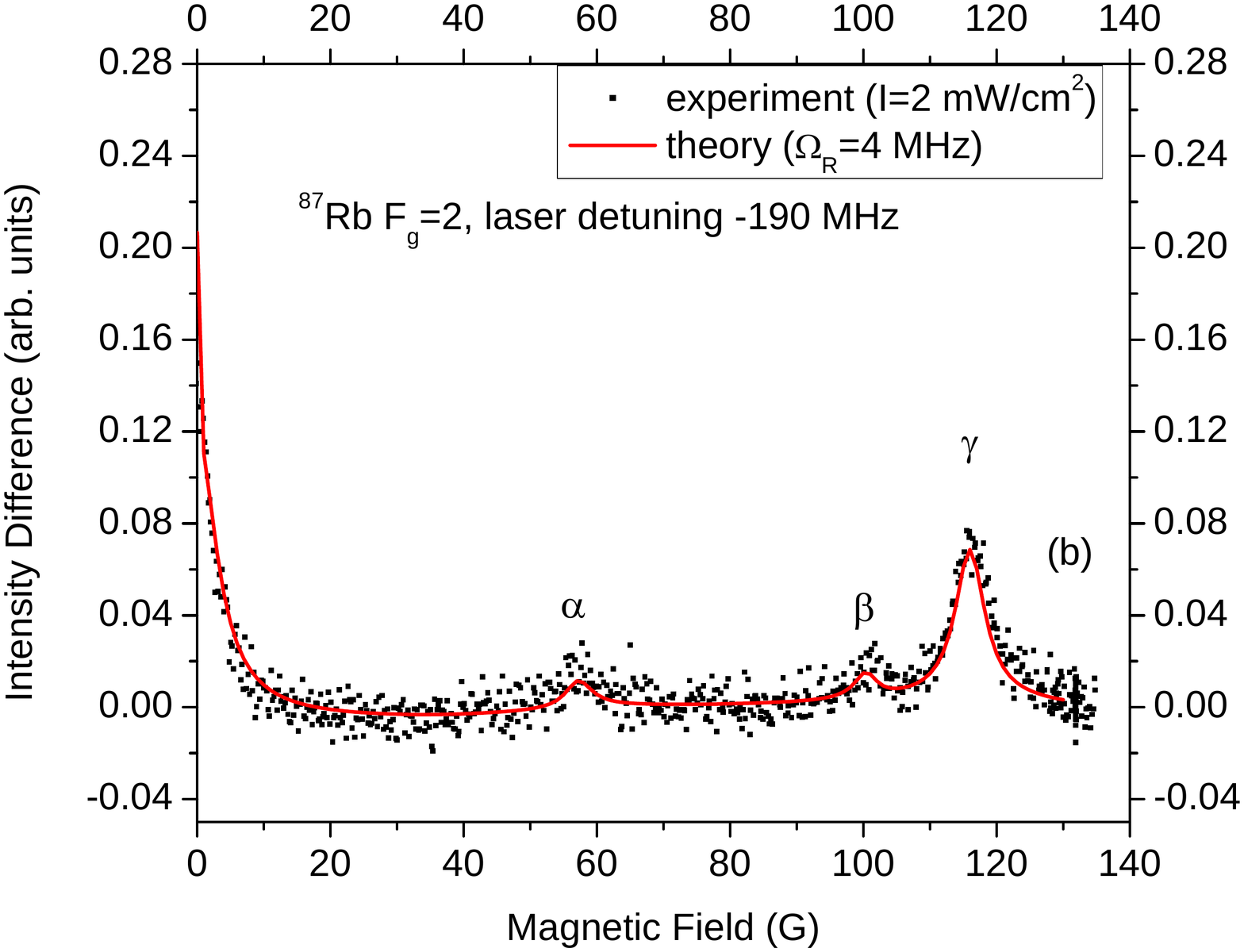}}
	\caption{\label{rb87-fg2} (Color online) Intensity difference ($I_{\perp}-I_{\parallel}$) versus magnetic field for the $F_g=2\longrightarrow F_e=1,2,3$ transitions of 
$^{87}$Rb with the laser detuned from the $5^2S_{1/2} (F_g=2) \longrightarrow 5^2P_{3/2}$ transition by (a) 36 MHz and (b) by $-190$ MHz.  Markers show the results of an experiment with the laser power density $I=2$ mW/cm$^2$, while the solid line shows the results of a calculation with Rabi frequency $\Omega_R=4$ MHz. 
}
\end{figure*}

Figure~\ref{rb87-fg1} and Fig.~\ref{rb87-fg2} show results for $^{87}$Rb for excitation from the $F_g=1$ and $F_g=2$ ground 
state levels, respectively. In this case, the detuning can dramatically affect the shape of the signals. As can be seen by comparing 
Fig.~\ref{rb87-fg1}(a) and Fig.~\ref{rb87-fg1}(b), the contrast of the peak that corresponds to level crossing $\beta$ increased by more than a 
factor of two when the laser was detuned by 36 MHz as compared to when it was detuned by $-190$ MHz. Similarly, when the 
excitation took place from the ground state with $F_g=2$, the peak that corresponds level crossing $\alpha$ was not visible when the laser was detuned by 36 MHz 
[Fig.~\ref{rb87-fg2}(a)], but appeared when the laser was detuned by $-190$ MHz [Fig.~\ref{rb87-fg2}(b)]. The contrast of the peak at level crossing $\gamma$ also increases by a factor of two depending on if the laser is tuned to a value that closely corresponds to the energy difference between the ground state and its level crossing [Fig.~\ref{rb87-fg2}(b)] or far away [Fig.~\ref{rb87-fg2}(a)]. Note that "close" here does not mean in magnetic field values, but in energy difference (see Fig.~\ref{fig:levels}.) Again, these peaks illustrate the power of using a detailed model, as the resonance labeled $\beta$ would not have appeared in the theoretical calculation when exciting from $F_g=1$ without taking into account sublevel mixing. 

\begin{figure}[htbp]
	\centering
		\resizebox{\columnwidth}{!}{\includegraphics{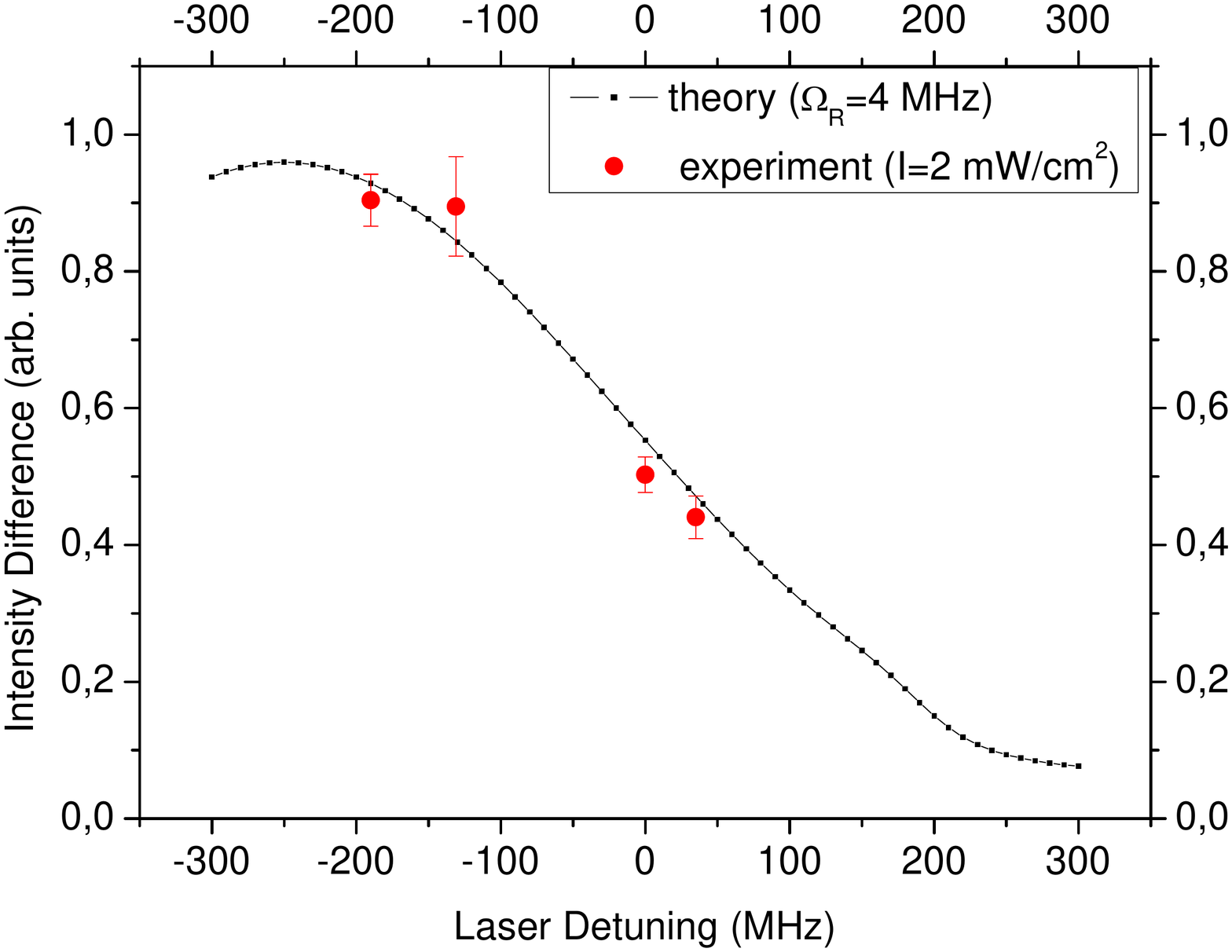}}
	\caption{\label{rb87-fg1-detuning} (Color online) Intensity difference ($I_{\perp}-I_{\parallel}$) versus laser detuning for fixed magnetic field ($B=57$ G) value for the transition from the ground state of $^{87}$Rb with $F_g=1$. 
}

\end{figure}

\begin{figure}[htbp]
	\centering
		\resizebox{\columnwidth}{!}{\includegraphics{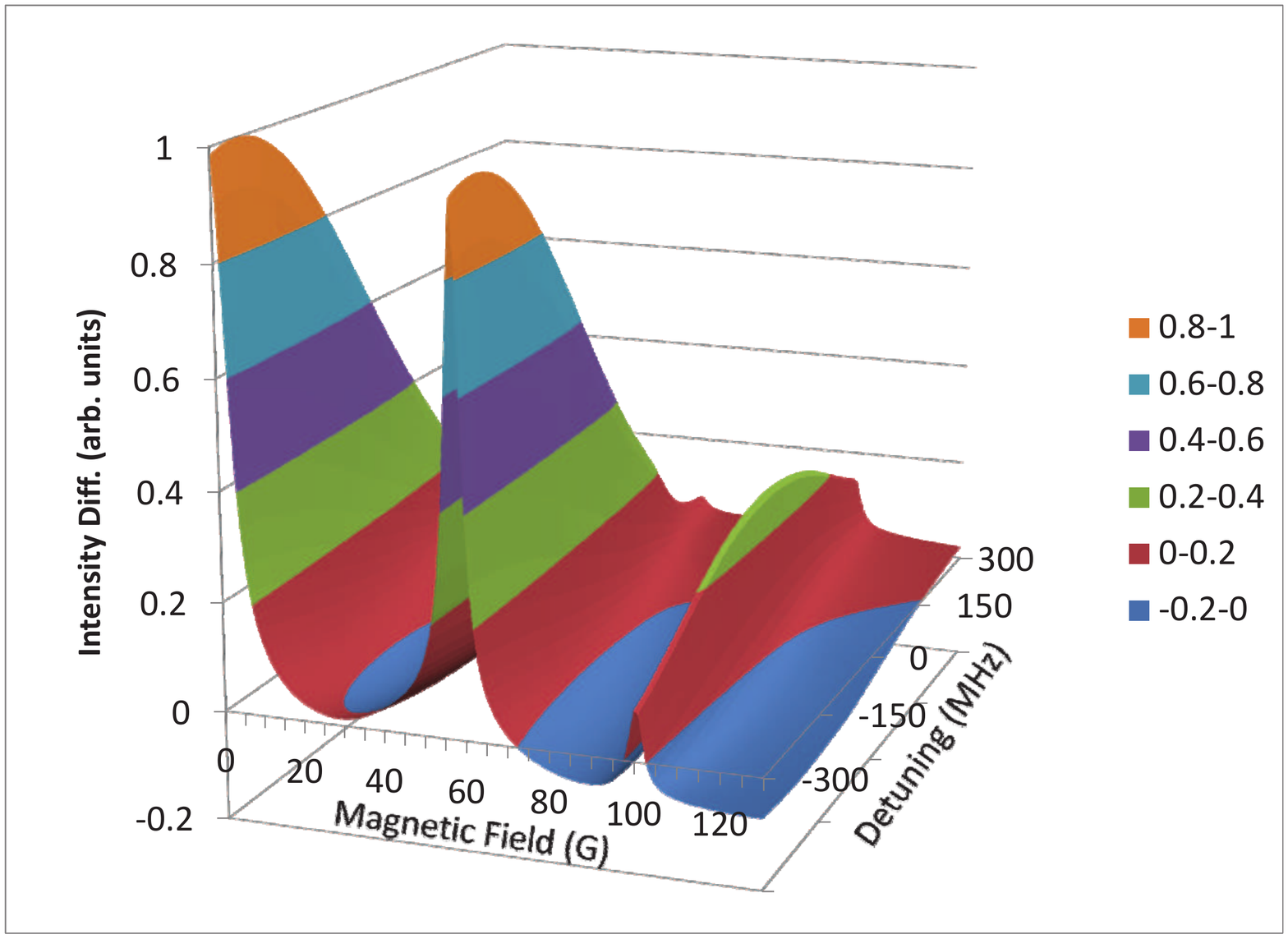}}
	\caption{\label{rb87-fg1-3D} (Color online) Intensity difference ($I_{\perp}-I_{\parallel}$) versus laser detuning and magnetic field for the transition from the ground state of $^{87}$Rb with $F_g=1$. 
}
\end{figure}

Figure~\ref{rb87-fg1-detuning} shows the intensity difference as a function of laser detuning when the magnetic field is fixed to a value of $B=57$ G and the excitation takes place from the ground state hyperfine level with $F_g=1$. One can clearly see that there is an optimal detuning for maximizing the intensity difference between the two polarization components. The curve has a half width at half maximum of around 250 MHz, comparable to the Doppler width of rubidium at room temperature.  The black squares connected by a line give the results of a theoretical calculation for a Rabi frequency of $\Omega_R=4$ MHz. The filled circles represent experimentally measured values for a laser power density of 2 mW/cm$^2$. The agreement is quite good. In order to optimize the experimental conditions under which the maximum contrast for a level-crossing peak could be observed, it can be useful to generate from the results of the calculations a three-dimensional plot of the intensity difference of the two orthogonally polarized components of the LIF as a function of laser detuning and magnetic field. Such a plot is shown in Fig.~\ref{rb87-fg1-3D} and from it one can easily determine the optimal laser detuning for maximizing the contrast of each level-crossing peak.  


As is well-known, the level-crossing resonances are manifestations of coherences in the excited state manifold of the atoms under study. These coherences can be visualized by plotting surfaces that represent the probability of finding the angular momentum of an atom pointing in a particular direction in space~\cite{Auzinsh:1997, Rochester:2001}. Such surfaces can be generated from the density matrices by summing over all $F$ states (see~\cite{Auzinsh:2006}), which yields the spatial distribution of $J$. Such surfaces have been included in Fig.~\ref{rb87-fg1} as an illustration of the information that can be gleaned from the theoretical model. The axes $I_{x}$ and $I_{y}$ refer to the states that give rise to the respectively polarized fluorescence when they decay. When two or more sublevels in the excited state that could be excited coherently ($\Delta m=\pm2$) are degenerate, the excited state becomes aligned, which manifests itself as a nonuniform angular momentum distribution in the $xy$-plane. In other words, an aligned state gives rise to an angular momentum distribution that has the $z$-axis as a second-order symmetry axis. In the absence of coherence the angular momentum spatial distribution is axially symmetric with respect to the $z$-axis. Aligned states occur when all sublevels cross at zero magnetic field and at the level-crossing points that can be excited coherently, such as $\alpha$ and $\beta$ in Fig.~\ref{rb87-fg1}. Far from the level-crossing points the angular momentum distributions become symmetric. The degree of asymmetry in the angular momentum distribution also reflects the strength of the coherence and the contrast of the respective resonance. 

\section{\label{Conclusion:level1}Conclusion}
 	The results presented in this work demonstrate two things: (1) the shapes of $\Delta m=\pm2$ level-crossing signals can be theoretically modeled with good accuracy over a broad range of magnetic field values and over at least an order of magnitude in laser power density; and (2) the detuning of the laser can dramatically influence the shape of the level-crossing curve, in particular the contrast of the peaks, when the hyperfine splitting of the excited state exceeds the Doppler broadening. This sensitivity to the laser detuning also provided a particularly stringent test of the theoretical model, as well as being interesting in their own right for the optimization of level-crossing studies. Furthermore, the density matrices computed with the theoretical model can shed light on the coherent processes associated with the excitation of the atoms, which is useful for computing fluorescence intensities for arbitrary directions and polarizations.  Although the atomic constants involved in the transitions of the $D_2$ line of rubidium are well known, a precise model of level-crossing signals can be necessary for extracting atomic constants in situations where the large number of level crossings washes out individual peaks or where the hyperfine splitting is small~\cite{Ney:1968}. The utility of such a model for the hyperfine constants $A$ for the $7,9,10D_{5/2}$ states of cesium from electric field level crossings was demonstrated, for example, in~\cite{Auzinsh:2007}.

\begin{acknowledgments}
We thank Ronald Rundans for help with the experiments. This work was supported by 
ERAF project Nr. 2010/0242/2DP/2.1.1.1.0/10/APIA/VIAA/036 and the Latvian State Research Programme Nr. 2010/10-4/VPP-2/1. 
M.~A. gratefully acknowledges support from the NATO Science for Peace project CBP.MD.SFPP. 983932, "Novel Magnetic Sensors and Techniques for Security Applications."

\end{acknowledgments}
\bibliography{rubidium}
\end{document}